\documentclass[a4paper,12pt]{article}
\usepackage{graphicx}               %
\usepackage{times}
\usepackage[dvips,unicode,colorlinks,linkcolor=blue,citecolor=blue,urlcolor=blue]{hyperref}
\begin{document}
\author{A. V. Savin$^{1,2}$ \and
        O. V. Gendelman$^{1,3}$}

\title{Heat conduction in 1D lattices with on-site potential}
\maketitle
\footnotetext[1]{Institute of Chemical Physics RAS, Kosygin str.4, Moscow, Russia}
\footnotetext[2]{asavin@center.chph.ras.ru}
\footnotetext[3]{ovgend@center.chph.ras.ru}

\tableofcontents
\begin{abstract}
The process of heat conduction in one-dimensional lattice with on-site potential is studied by
means of numerical simulation. Using discrete Frenkel-Kontorova, $\phi$--4 and sinh-Gordon we
demonstrate that contrary to previously expressed opinions the sole anharmonicity of the
on-site potential is insufficient to ensure the normal heat conductivity in these systems. The
character of the heat conduction is determined by the spectrum of nonlinear excitations
peculiar for every given model and therefore depends on the concrete potential shape and
temperature of the lattice. The reason is that the peculiarities of the nonlinear excitations
and their interactions prescribe the energy scattering mechanism in each model. For models
sin-Gordon and $\phi$--4 phonons are scattered at thermalized lattice of topological solitons;
for sinh-Gordon and $\phi$--4 - models the phonons are scattered at localized high-frequency
breathers (in the case of $\phi$--4 the scattering mechanism switches with the growth of the
temperature).
\end{abstract}

\section{Introduction}
Heat conductivity of 1D lattices is well known classical problem related to microscopic
foundation of Fourier law.  The problem started from the famous work of Fermi, Pasta and Ulam
\cite{p1}, where abnormal process of heat transfer has been detected in the first time.
Non-integrability of the system is necessary condition for normal heat conductivity. As it was
demonstrated recently for the FPU lattice \cite{p2,p3,p4}, disordered harmonic chain
\cite{p4a,p4b,p4c}, diatomic 1D gas of colliding particles \cite{p4d,p4e,p4f} and the diatomic
Toda lattice \cite{p5}, the non-integrability is not sufficient in order to get normal heat
conductivity. It leads to linear distribution of the temperature along the chain, but the value
of heat flux is proportional to $1/N^{\alpha}$, where $0<\alpha<1$ and $N$ is the number of
particles in the chain. Thus, the coefficient of the heat conductivity diverges in
thermodynamical limit $N\rightarrow\infty$. Analytical estimations \cite{p4} have demonstrated
that any chain possessing acoustic phonon branch should have infinite heat conductivity in the
limit of low temperatures.

From the other side, there are some artificial systems with on-site potential having normal
heat conductivity \cite{p7,p8}. Heat conductivity of Frenkel-Kontorova chain in the first time
was considered in paper \cite{p12}. Finite heat conductivity for certain set of parameters was
obtained for Frenkel-Kontorova chain  \cite{p9}, for the chain with sinh-Gordon on-site
potential \cite{p10} and for the chain with  $\phi^4$ on-site potential \cite{p10a,p10b}. These
models are not invariant with respect to translation and the momentum is not conserved. It was
supposed that the on-site potential is extremely significant for normal heat conduction
\cite{p10a} and that the anharmonicity of the on-site potential is sufficient to ensure the
validity of Fourier law \cite{p11}. Detailed review of the problem is presented in recent paper
\cite{p15a}.

Peculiarities of the heat conduction of the Frenkel-Kontorova model for complete set of
parameters and temperature of the system are not known.
The chains with zero average pressure were demonstrated to have normal heat conductivity
\cite{p13,p14,p15}. In papers \cite{p14,p15} the transition from abnormal to normal heat
conductivity has been detected at certain temperature. Paper \cite{p15a} contains detailed
review of the heat conductivity peculiarities in 1D molecular systems.

There are no detailed investigations similar to mentioned above and concerning the properties
of the chains with on-site potential in the whole temperature range. As it was mentioned above,
there exists certain incompleteness of the knowledge concerning even the most popular and
paradigmatic discrete Frenkel-Kontorova chain. This lattice is of special interest as its
continuous counterpart is famous sin-Gordon system (having, of course, divergent heat
conductivity).  The transition between two regimes with the growth of temperature is expected
for discrete system; however it may be dependent also on other parameters of the lattice

The question of special interest is also the mechanism of heat flow scattering which gives rise
to finite heat conductivity. For the chain with periodic nearest-neighbor interaction it was
demonstrated \cite{p14,p15} that the transition  to normal heat conductivity corresponds to
abrupt growth of concentration of rotation solitons (rotobreathers), demonstrating certain
similarity with phase transition. Namely, the region of the transition temperature corresponds
to maximum region of the heat capacity of the lattice. Similarly, it is reasonable to suppose
that every lattice with finite heat conductivity has its peculiar mechanism of scattering the
heat flow.

The paper is devoted to the detailed simulation of the discrete lattices with on-site
nonlinearity and quadratic potential of nearest-neighbor interaction and investigation of
their heat conductivity. The lattices are Frenkel-Kontorova, sinh-Gordon, and discrete
$\phi$-4. For every case the dependence of the heat conductivity on the temperature and
parameters of the lattice will be explored and concrete elementary excitation responsible
for the change of regimes will be revealed.

\section{Description of the model}

Let us consider one-dimensional atomic chain arranged  along $x$ axis. All particles are of
equal mass $M$, and the nearest-neighbor interaction is described by harmonic potential having
rigidity $K$. Then the Hamiltonian of the lattice will take a form
       \begin{equation}
    {\cal H}=\sum_n\{ \frac12 M\dot{x}_n^2+ \frac12 K(x_{n+1}-x_n)^2+U(x_n)\},
       \label{f1}
       \end{equation}
where the dot denotes the differentiation with respect to  time $t$, $x_n$ is the displacement
of the  $n$-th particle from its equilibrium position and $U(x)$ is the on-site potential.

The dimensionless variables are introduced as $u_n=2\pi x_n/a$ ($a$ is the  equilibrium
distance between the particles) for the displacement, $\tau=t\sqrt{K/M}$ for the time and
$H=4\pi^2{\cal H}/Ka^2$ for the energy. Hamiltonian (\ref{f1}) takes the form
       \begin{equation}
       H=\sum_n\{ \frac12{u'_n}^2+ \frac12 (u_{n+1}-u_n)^2+V(u_n)\},
       \label{f2}
       \end{equation}
where the stroke denotes the differentiation with respect to the dimensionless time $\tau$ and
the dimensionless on-site potential is introduced as $V(u_n)=4\pi^2 U(au_n/2\pi)/Ka^2$. Natural
definition for the dimensionless temperature is  $T=4\pi^2 k_B\Theta/Ka^2$, where $k_B$ is
Boltzmann constant and $\Theta$ is the temperature in common units.

We are going to consider four widely used models for on-site potential: harmonic potential
       \begin{equation}
       V(u)=\frac12\omega_0^2 u^2;
       \label{f3}
       \end{equation}
 sin-Gorgon potential
       \begin{equation}
       V(u)=\epsilon [1+\cos(u)];
       \label{f4}
       \end{equation}
 $\phi$-4 potential
       \begin{equation}
       V(u)=2\epsilon [(u/\pi)^2-1]^2
       \label{f5}
       \end{equation}
and sinh-Gordon potential
       \begin{equation}
       V(u)=\omega_0^2 [\cosh(u)-1].
       \label{f6}
       \end{equation}
Parameter $\epsilon>0$ determines the value of potential barrier between  neighboring wells and
its inverse $g=1/{\epsilon}$ characterizes the cooperativeness of the system. Potentials
(\ref{f4}) and (\ref{f5}) have the same distance between neighboring wells equal to 2$\pi$ and
equal value of the potential barrier 2$\epsilon$. Parameter {$\omega_0$} in  (\ref{f3}) and
(\ref{f6}) corresponds to minimal frequency of harmonic vibrations of the lattice.

\section{Methods for computation of the heat conduction coefficient}

\begin{figure}[t]
\begin{center}
\includegraphics[angle=0, width=0.85\textwidth]{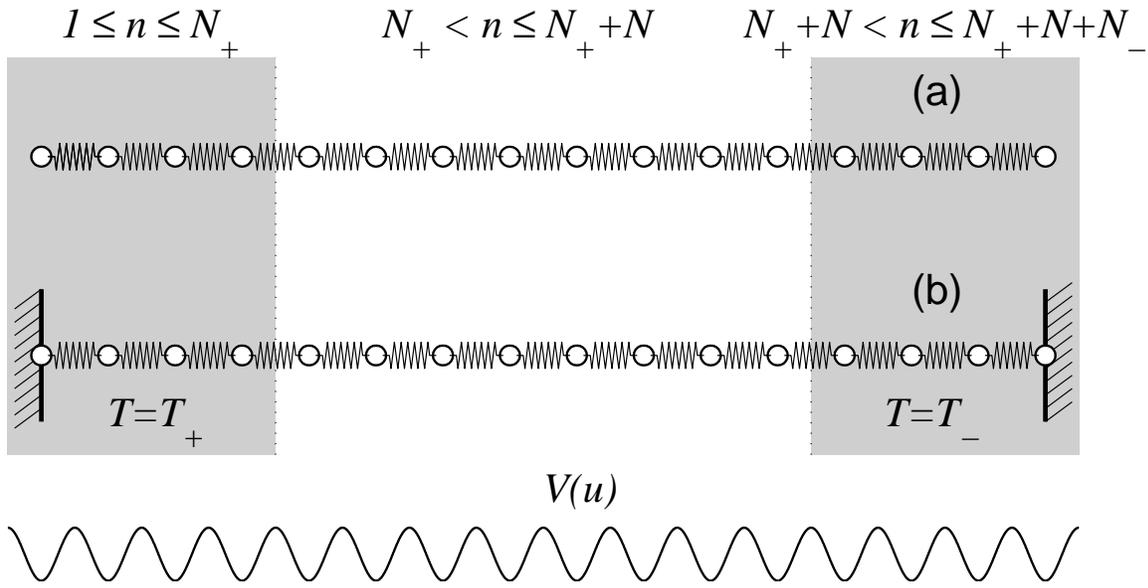}
\end{center}
\caption{\label{fig1}\protect\small
       Model of the chain of $N_++N+N_-$ particles with
       left $N_+$ particles attached to
       $T=T_+$ thermostat and right $N_-$ particles attached to
       $T=T_-$ thermostat. Boundary conditions correspond to free (a)
       and fixed (b) end particles. The potential $V(u)$ corresponds
       to discrete Frenkel - Kontorova model.
        }
\end{figure}

The goal is to simulate the process of heat conduction in finite chain containing N particles.
For this sake the left side of the chain ($n\le 0$) has to be connected to a thermostat with
temperature  $T_+$, and the right side ($n>N$) -- to a thermostat with temperature $T_-$
($T_+>T_-$). For the sake of the simulation we consider the chain of $N_-+N+N_+$ particles,
where the first $N_+$ particles are attached to the thermostat $T_+$, an the last $N_-$
particles -- to the thermostat $T_-$ (Fig. \ref{fig1}). The potential of the nearest-neighbor
interaction is harmonic, therefore the equilibrium length of the chain does not depend on the
temperature. It implies that the boundary conditions at the ends of the lattice has no noticeable
effect on the
process of the heat conduction and both the conditions of free  (Fig.\ref{fig1} (a)) and fixed
(Fig. \ref{fig1} (b)) may be used. Numerical simulations with $N_\pm=40$ has demonstrated the
absence of the effect depending on concrete choice of the boundary conditions. We'll use the
condition of free ends with $N_\pm=40$ for all simulations.

Majority of papers devoted to the topic of the heat conduction \cite{p2,p3,p9,p10a} uses
deterministic Nose - Hoover thermostat \cite{p16} with $N_+=N_-=1$. However this thermostat has
been designed for the description of the equilibrium thermalized system and is not universally
suitable for the description of non-equilibrium processes. Therefore our choice is well-known
stochastic Langevin thermostat. Detailed comparison of these two thermostats is presented in
Appendix \ref{pr3}.

Let us consider the chain with free ends ($1<n<N+N_++N_-$) with $N_\pm$ particles at both ends
attached to Langevin thermostats. The dynamics of the system is described by equations
        \begin{eqnarray}
        u''_n&=&u_{n+1}-u_n-F(u_n)-\gamma u'_n+\xi_n^+,
        \nonumber \\
             n&=&1, \nonumber \\
        u''_n&=&u_{n+1}-2u_n+u_{n-1}-F(u_n)-\gamma u'_n+\xi_n^+,
        \nonumber \\
        n&=&2,...,N_+,\nonumber \\
        u''_n&=&u_{n+1}-2u_n+u_{n-1}-F(u_n),
        \label{f7} \\
        n&=&N_++1,...,N_++N, \nonumber \\
        u''_n&=&u_{n+1}-2u_n+u_{n-1}-F(u_n)-\gamma u'_n+\xi_n^-,
        \nonumber \\
        n&=&N_++N+1,...,N_++N+N_--1,
        \nonumber \\
        u''_n&=& u_{n-1}-u_n-F(u_n)-\gamma u'_n+\xi_n^-,
        \nonumber \\
        n&=&N_++N+N_-, \nonumber
        \end{eqnarray}
where $F(u)=dU(u)/du$, the damping coefficient $\gamma=1/\tau_r$, $\tau_r$ is the
characteristic relaxation time of the particles attached to the thermostat, $\xi_n^\pm$ is the
random external force corresponding to Gaussian white noise normalized as
        \begin{eqnarray}
        \langle\xi_n^\pm(\tau)\rangle=
        \langle\xi_n^\pm(\tau_1)\xi_k^\mp(\tau_2)\rangle=0,\nonumber \\
        \langle\xi_n^\pm(\tau_1)\xi_k^\pm(\tau_2)\rangle
        =2\gamma T_\pm\delta_{nk}\delta(\tau_2-\tau_1). \nonumber
        \end{eqnarray}
Details of numeric realization of the Langevin thermostat and random forces are presented in
Appendix \ref{pr1}.

At every moment the dimensionless temperature of the $n$-th particle
$t_n(\tau)={u'_n}^2(\tau)$. In order to determine the value of the local heat flux $j_n$ the
energy distribution among the particles of the chain is considered:
       \begin{equation}
       h_n=\frac12\left[\frac12({u'_n}^2+{u'_{n+1}}^2)+V(u_n)+V(u_{n+1})
       +(u_{n+1}-u_n)^2\right].
       \label{f8}
       \end{equation}
By differentiating equation (\ref{f8}) with respect to time $\tau$ we get
       $$
       h'_n=\frac12\left\{ u'_n[u''_n+F(u_n)]
           +u'_{n+1}[u''_{n+1}+F(u_{n+1})]\right\}
           +(u_{n+1}-u_n)(u'_{n+1}-u'_n).
       $$
With account of (\ref{f7}) we obtain
      \begin{equation}
      h'_n=\frac12\left[u'_{n+1}(u_{n+2}-u_n)-u'_n(u_{n+1}-u_{n-1})\right].
      \label{f9}
      \end{equation}
Taking  into account the continuity condition $h'_n=j_n-j_{n-1}$ we get the expression for the
energy flux:  $j_n=-u'_n(u_{n+1}-u_{n-1})/2$.

System of equations (\ref{f7}) has been integrated numerically. We  used the values of
$\gamma=0.1$, $N_\pm=40$, $N=10$, 20, 40, 80, 160, 320, 640 and initial conditions
corresponding to the ground state of the chain. After the time $\tau=10^5$ has elapsed, the end
particles achieved thermal equilibrium with the thermostat and stationary heat flux has been
formed. Afterwards the dynamics of system (\ref{f7}) has been simulated at the time scale of
order $\tau=10^7$. The average temperatures of the particles
       \begin{equation}
       T_n=\langle t_n(\tau)\rangle_\tau=\lim_{\tau\rightarrow\infty}
       \frac1\tau\int_0^\tau {u'_n}^2(s)ds.
       \label{f10}
       \end{equation}
and average values of the heat flux
       \begin{equation}
       J_n=\langle j_n(\tau)\rangle_\tau=\lim_{\tau\rightarrow\infty}
       \frac1\tau\int_0^\tau j_n(s)ds.
       \label{f11}
       \end{equation}
were computed for the fragment of the chain between the thermostats.

\begin{figure}[tp]
\begin{center}
\includegraphics[angle=0, width=0.85\textwidth]{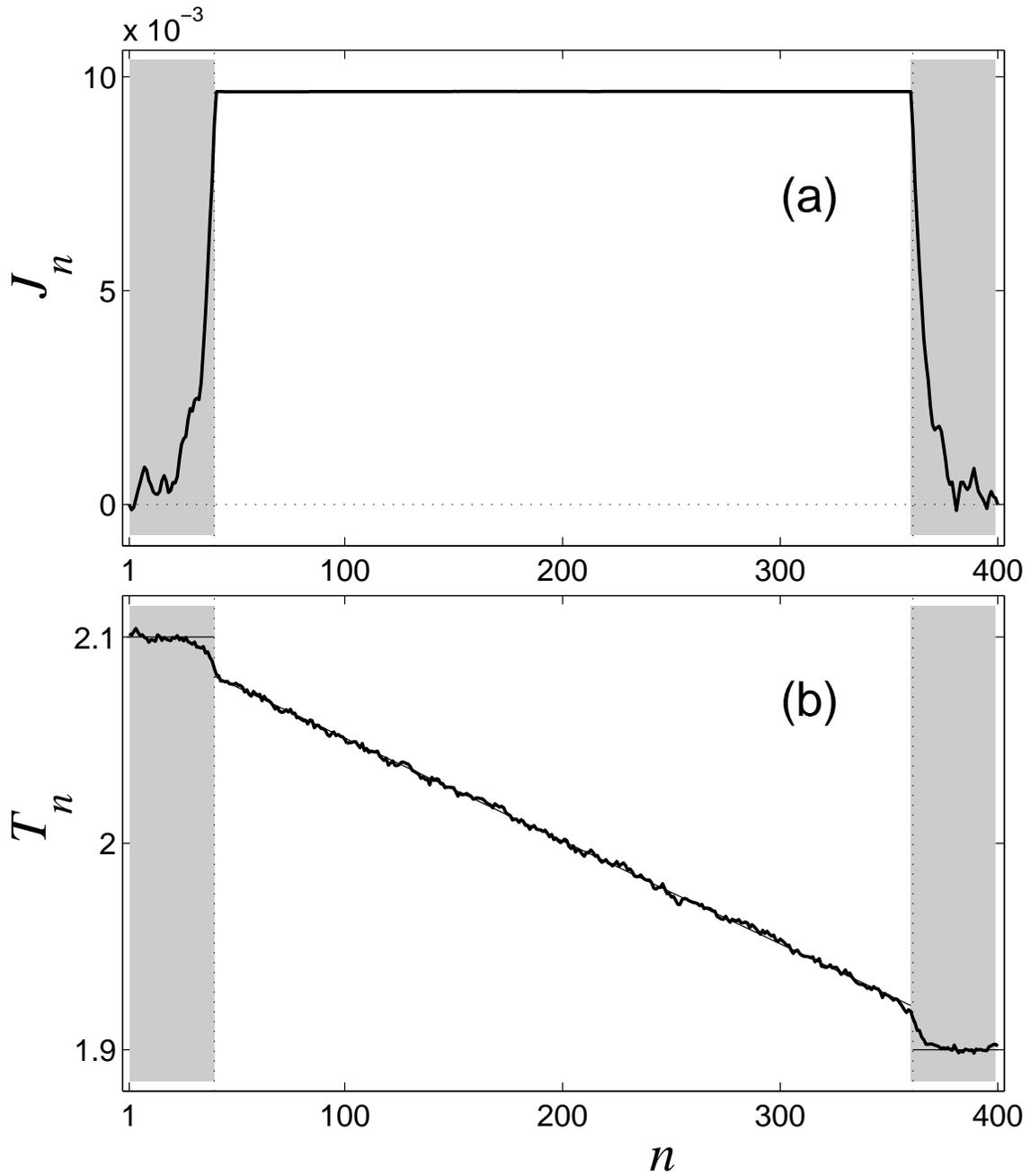}
\end{center}
\caption{\label{fig2}\protect \small
       Distribution of the local heat flux $J_n$ (a) and local temperature $T_n$
       (b) in the chain with periodic on-site potential (\ref{f4}),
       $\epsilon=1$, $N=320$, $N_\pm=40$, $T_+=2.1$, $T_-=1.9$.
       Time of averaging $\tau=10^7$. Fragments of the chain interacting with the
       thermostats are embedded in gray.
        }
\end{figure}

If the temperature gradient $\Delta T=T_{+}-T_{-}$ is small, this  method allows avoiding the
temperature jumps at the ends of the free fragment of the chain \cite{p18}. Characteristic
distributions of the heat flux $J_n$ and local temperature $T_n$ are demonstrated at Fig.
\ref{fig2}. At the inner fragment of the chain $N_+<n\le N_++N$ the heat flux is constant
$(J_n=J)$ and the temperature profile is linear. The coefficient of the heat conductivity is
determined using the information concerning the inner fragment of the chain:
       \begin{equation}
       \kappa(N)=J(N-1)/(T_{N_++1}-T_{N_++N}).
       \label{f12}
       \end{equation}
If the function $T=\alpha n+b$ is the best linear approximation for $T=T_n$ data at the inner
fragment of the chain $N_+<n\le N_++N$, then the value $\kappa (N)$ may be calculated with
better accuracy by taking $\kappa (N)=J\alpha$.

Limit value
       \begin{equation}
       \kappa=\lim_{N\rightarrow\infty}\kappa(N)
       \label{f13}
       \end{equation}
will correspond  to the coefficient of the heat conductivity at temperature
$T=(T_{+}+T_{-})/2$. The question regarding the finiteness of the heat conductivity is reduced
to the existence of finite limit (\ref{f13}). I the sequence $\kappa(N)$ diverges as
$N\rightarrow\infty$ then the chain has infinite heat conductivity at this value of the
temperature.

Alternative way to compute the heat conductivity $\kappa$ is related to well-known  Green-Kubo
formula \cite{p19}:
       \begin{equation}
       \kappa =\lim_{\tau\rightarrow\infty}\int_{0}^\tau
       \lim_{N\rightarrow\infty}\frac{1}{NT^2}\langle {\bf J}(s){\bf J}(0)\rangle ds~,
       \label{f14}
       \end{equation}
where $N$ is already the number of particles in the chain with periodic boundary conditions,

$$
{\bf J(\tau)}=\sum_{n=1}^N j_n(\tau)
$$
is the general heat flux and the averaging $\langle\cdot\rangle$ is performed over  all
thermalized states of the chain. Consequently, the finiteness of the heat conductivity is
related to the convergence of the integral
       \begin{equation}
       \int_0^\infty C(\tau)d\tau, \label{f15}
       \end{equation}
e.g. to the evaluation of the descending rate of the function
       $$
       C(\tau)=\lim_{N\rightarrow\infty}\frac{1}{NT^2}\langle {\bf J}(\tau){\bf J}(0)\rangle.
       $$

Numerically the above autocorrelation function may be found only for finite chain
       \begin{equation}
       C_N(\tau)=\frac{1}{NT^2}\langle {\bf J}(s){\bf J}(s-\tau)\rangle_s. \label{f16}
       \end{equation}
For large enough values of $N$ the correlation function $C_N(\tau)$ is believed to approximate
the function $C(t)$ with acceptable accuracy. In order to get stable results the value $N=4000$
is usually sufficient. More details concerning the computation of the autocorrelation function
are presented in Appendix \ref{pr2}.

The methods for computing the heat conductivity coefficient (\ref{f13}), (\ref{f14}) are
complementary and allow mutual verification of the results.

\section{Heat conductivity of the chain with harmonic on-site potential}

\begin{figure}[tp]
\begin{center}
\includegraphics[angle=0, width=0.85\textwidth]{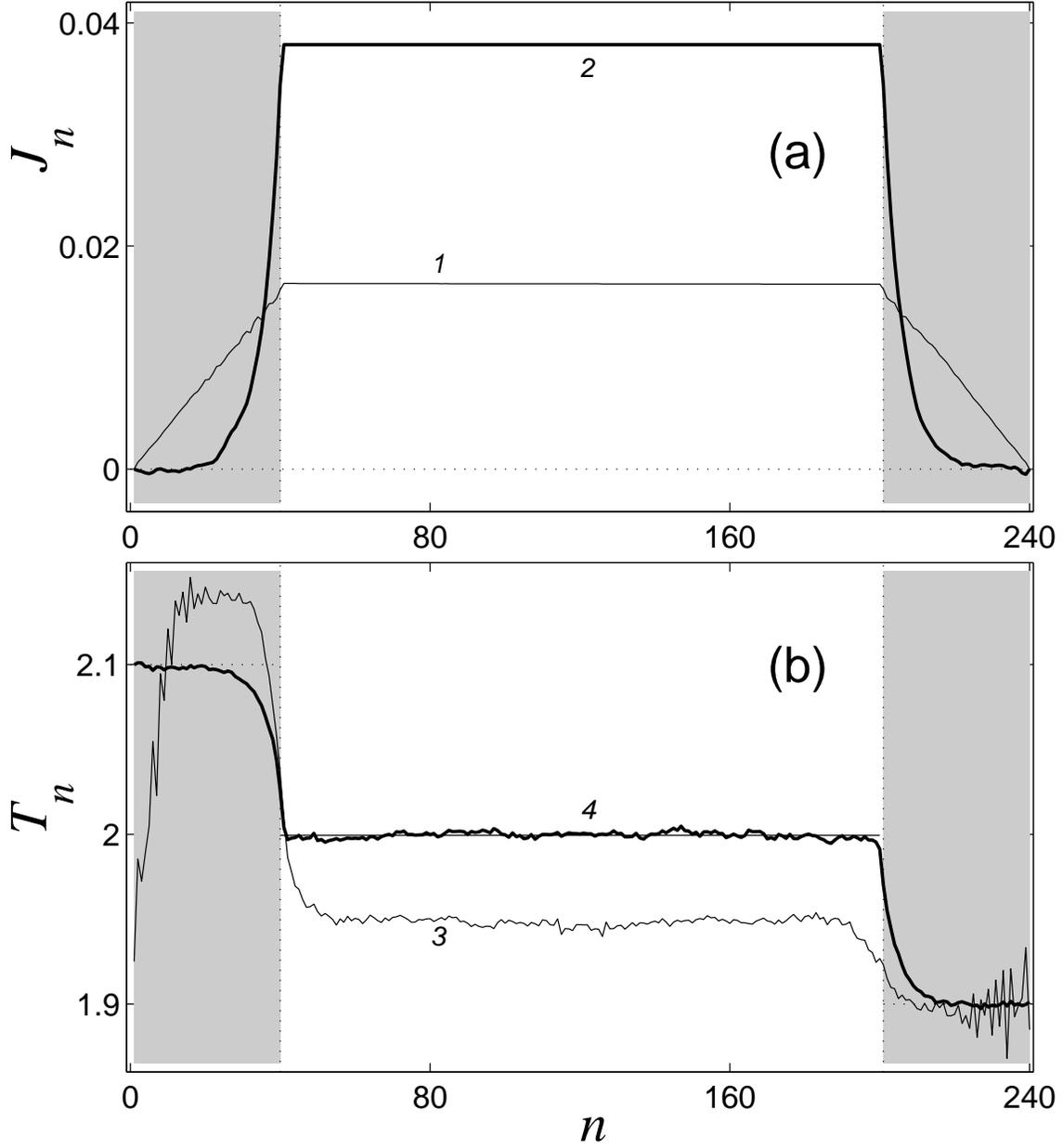}
\end{center}
\caption{\label{fig3}\protect \small
       Distribution of local heat flux $J_n$ (a) and local temperature $T_n$ (b) in the chain
       with harmonic on-site potential (\ref{f3}),
       $\omega_0=1$, $N=160$, $N_\pm=40$, $T_+=2.1$, $T_-=1.9$,
       averaging time $\tau=10^7$. The fragments of the chain interaction with the
       thermostats are embedded in grey. Thin lines (1, 3) are obtained by using the
       Nose-Hoover thermostat with $\tau_r=1$, and thick (2, 4) -- by using the
       Langevine thermostat with $\tau_r=10$.
        }
\end{figure}

The chain with harmonic on-site potential (\ref{f3}) is described by linear equations  and
therefore is completely integrable. The energy transport is performed by non-interacting phonon
modes. The heat flux $J$ does not depend on the chain length $N$, but only on the temperature
difference $\Delta T$. Linear thermal profile is not formed. At the inner part of the chain the
temperature is nearly constant $T_n=(T_++T_-)/2$ (Fig. \ref{fig3}). Therefore according to
(\ref{f12}), the heat conductivity coefficient diverges. Correspondingly, the average
correlation function $C(\tau)$ is constant and integral  (\ref{f15}) diverges.

\section{Heat conduction of the chain with periodic on-site potential}

Characteristic  features of dynamics of the chain with periodic on-site potential (\ref{f4})
depend on the values of the temperature. As the temperature is small $T\ll\epsilon$, the
on-site potential may be approximated by harmonic single-well potential (\ref{f3}) with
$\omega_0=\sqrt{\epsilon}$. The heat transport is governed by weakly interacting phonons. At
the temperature $T\sim\epsilon$ the chaotic superlattice of topological solitons is formed and
the transport properties change drastically. At very high temperatures $T\gg\epsilon$ the chain
is effectively detached from the site and again weakly interacting phonons govern the heat
transport. Therefore it is reasonable to investigate the dependence of the heat conductivity on
the reduced temperature $\tilde{T}=T/\epsilon$.

The behavior of  the chain also depends on the cooperativeness parameter $g=1/\epsilon$. The
more the cooperativeness, the less is the density of the soliton superlattice and the phonon
scattering effects are less significant.  The limit $g\rightarrow\infty$ ($\epsilon\rightarrow
0$) corresponds to completely integrable continuum sin-Gordon equation.

Generally, three limits of discrete  Frenkel-Kontorova system correspond to completely
integrable systems: at $\tilde{T}\rightarrow 0$ the system reduces to the harmonic chain with
harmonic on-site potential; at $\tilde{T}\rightarrow\infty$ -- to isolated harmonic chain; at
$g\rightarrow\infty$ -- to continuum sin-Gordon equation. All these limit systems have
diverging heat conductivity. The behavior of the system in the vicinity of these limits is a
natural question to be addressed.

\begin{figure}[tp]
\begin{center}
\includegraphics[angle=0, width=0.85\textwidth]{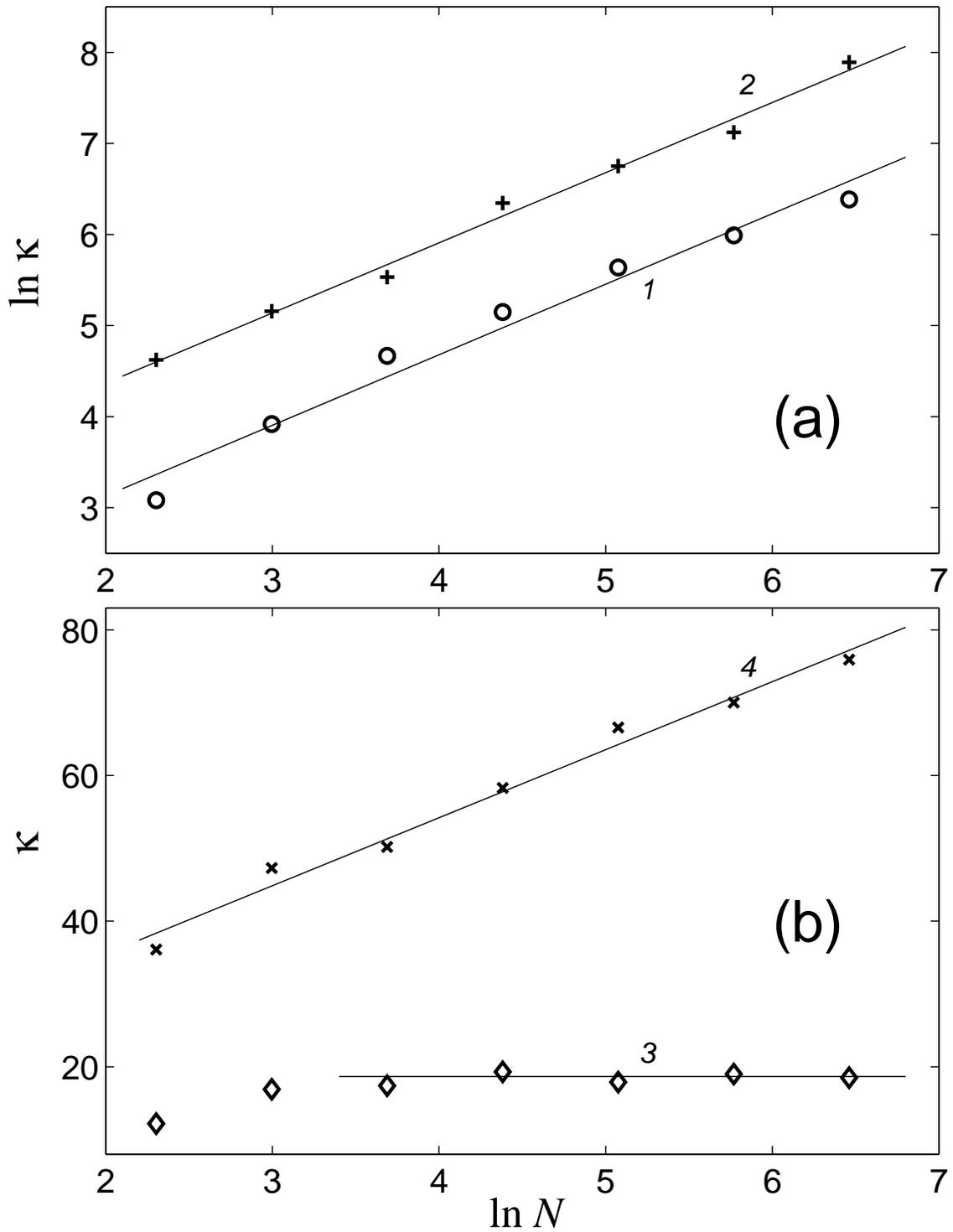}
\end{center}
\caption{\label{fig4}\protect \small
       Dependence of the logarithm of the heat conductivity coefficient
       $\ln\kappa (N)$ (a) and $\kappa (N)$ (b) on the logarithm of the
       inner fragment length $\ln N$ ($N_+=N_-=40$) for the chain with
       periodic on-site potential (\ref{f4}), $\epsilon=1$, $T=0.2$ (markers 1),
       $T=200$ (markers 2), $T=3$ (markers 3) and $T=20$ (markers 4).
       The markers denote the computed values and the lines correspond to the
       best linear approximations.
        }
\end{figure}

Let us start from $g=1$ $(\epsilon=1$, $\tilde{T}=T)$ and investigate the sequence $\kappa (N)$
as $N$ grows $(N=10,$ 20, 40, 80, 160, 320, 640) and different values of $T$. As it is may be
suggested from Fig. \ref{fig4} at small ($T=0.2$) and large ($T=200$) temperatures the heat
conductivity coefficient $\kappa (N)$ grows as $N^\alpha$, at $T=20$ -- as $\ln N$, and at
$T=3$ converges to finite value $\kappa=18.5$. Therefore it may be concluded that at $T=3$ the
chain has finite heat conductivity. The data related to the other values of the temperature
does not allow to draw any conclusions about the behavior of the heat conductivity at larger
values of $N$. Generally speaking, it may happen that for longer chains $\kappa(N)$ will attain
certain finite value. Computational tools we use do not allow to investigate higher values of
$N$. Stil, it is possible to get additional information from the behavior of the
autocorrelation function $C(\tau)$ at $\tau\rightarrow\infty$.

\begin{figure}[p]
\begin{center}
\includegraphics[angle=0, width=0.85\textwidth]{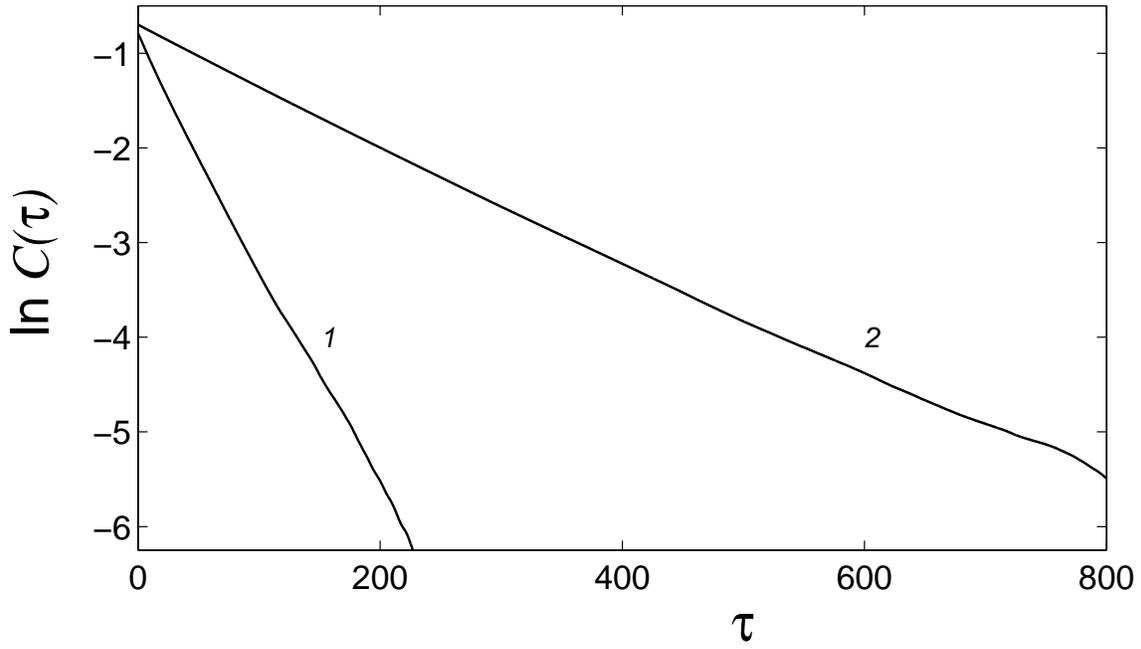}
\end{center}
\caption{\label{fig5}\protect \small
       Exponential decrease of the autocorrelation function $C(\tau)$
       in the chain with periodic on-site potential (\ref{f4}), $\epsilon=1$,
       $T=3$ (curve 1) and $T=20$ (curve 2) (semilogarythmic coordinates).
               }
\end{figure}
\begin{figure}[p]
\begin{center}
\includegraphics[angle=0, width=0.85\textwidth]{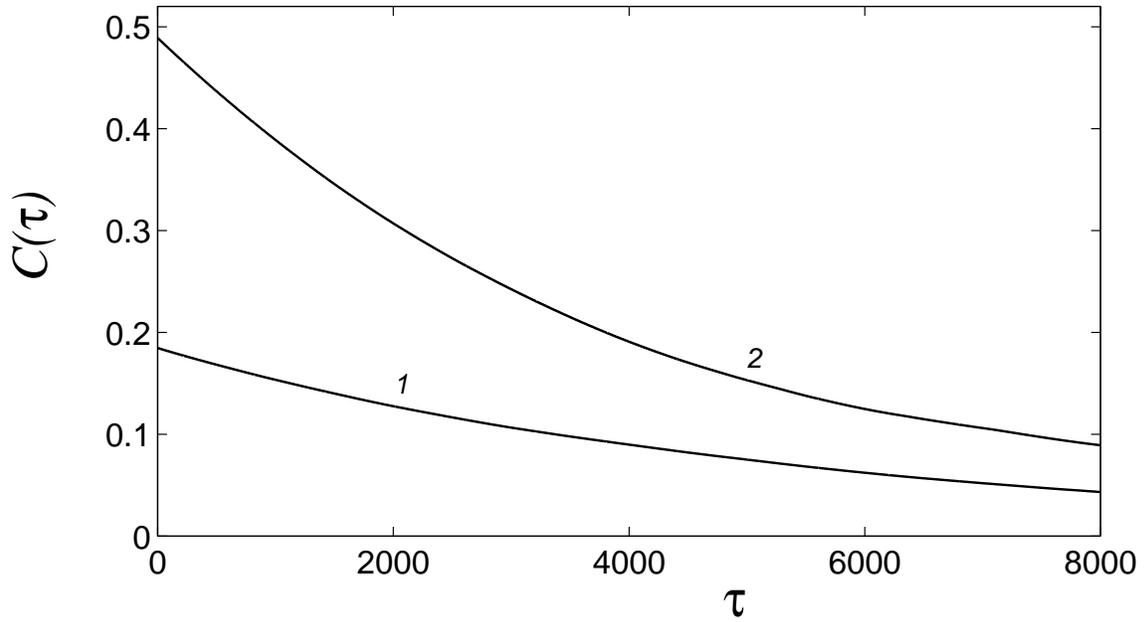}
\end{center}
\caption{\label{fig6}\protect \small
       Autocorrelation functions $C(\tau)$
       in the chain with periodic on-site potential (\ref{f4}), $\epsilon=1$,
       $T=0.2$ (curve 1) and $T=200$ (curve 2).
        }
\end{figure}
\begin{figure}[t]
\begin{center}
\includegraphics[angle=0, width=0.85\textwidth]{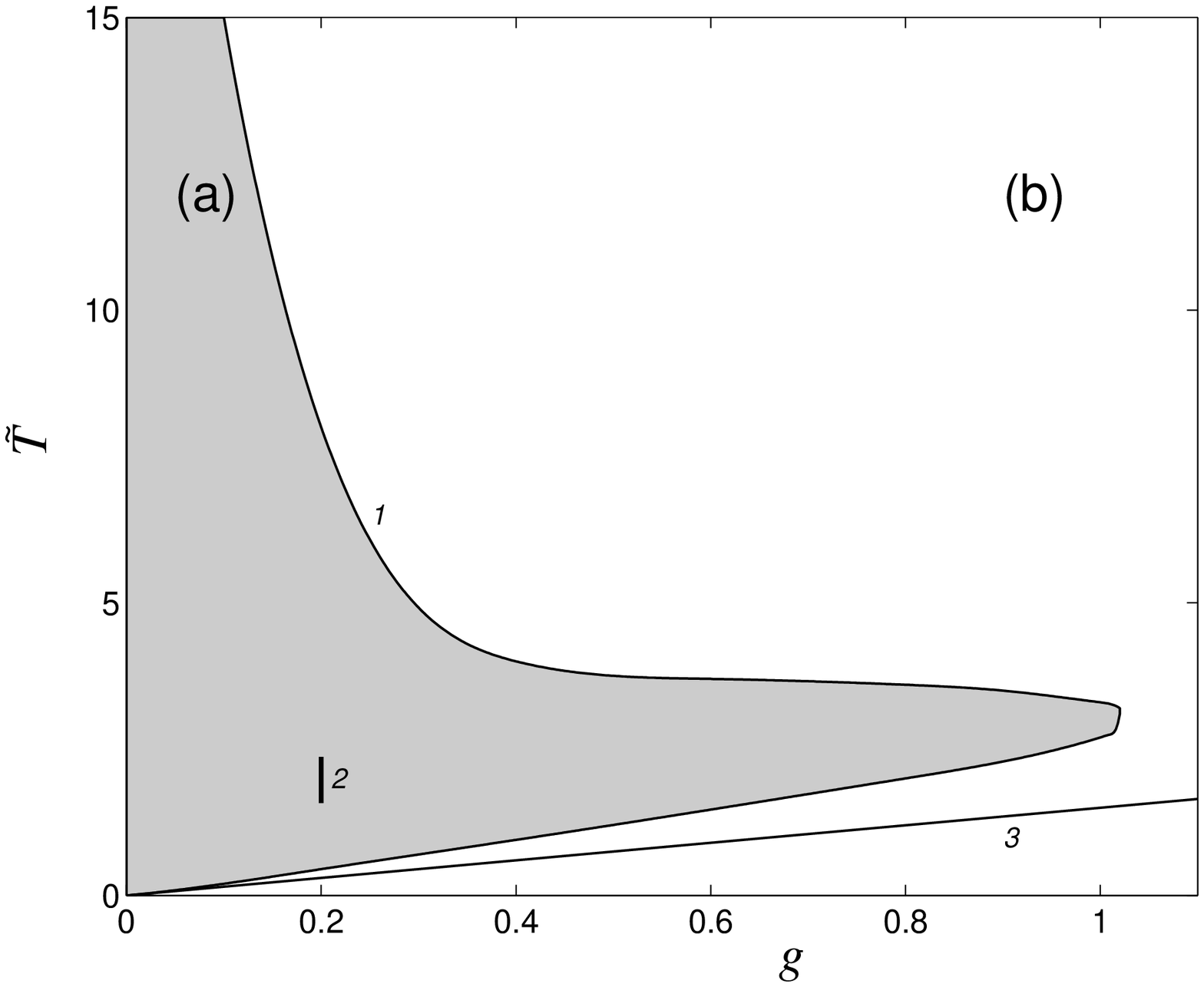}
\end{center}
\caption{\label{fig7}\protect \small
       The zone in the space of parameters $(g,\tilde{T})$, where for
       finite chains of length $N\le 640$ with on-site potential
       (\ref{f4}) he heat conductivity converges (a, grey zone) and
       diverges (b, white zone). Curve 1 divides these two zones. Interval 2 corresponds
       to the parameters used in \cite{p9}. For finite chains $(N\le 640)$
       with on-site potential $\phi$--4 (\ref{f5}) finite heat conductivity
       is detected only above line 3.
        }
\end{figure}

Numerical simulation demonstrates that for $T=3$ the autocorrelation function decreases
exponentially (Fig. \ref{fig5}, curve 1). Integral (\ref{f15}) converges and the Green-Kubo
formula (\ref{f14}) gives $\kappa=17.5$, in good correspondence with $\kappa=18.5$ obtained
from direct simulation of the heat flux. At $T=20$ the autocorrelation function at time scale
$0\le \tau\le 800$ also decreases exponentially (Fig. \ref{fig5}, curve 2). If this trend will
persist also for $\tau>800$, the Green-Kubo formula will give $\kappa=77.4$. It is reasonable
to compare this value with the result for $\kappa(N)$ presented at Fig. \ref{fig4} (curve 4).
Maximum value of $\kappa(640)=75.9$ and no trend towards any finite limit of $\kappa(N)$ may be
detected. Therefore the likely result is divergence. In order to verify this result the
simulation for larger values of $N$ (1280, 2560, 5120, 10240) is required, which is beyond our
computational possibilities.

The problem for $T=0.2$ and $T=200$ is even more difficult. The autocorrelation function  is
presented at Fig. \ref{fig6}. The decrease of the function is very slow and no unambiguous
conclusion concerning its character may be drawn out. While extrapolating $c(\tau)$ for
$\tau>8000$ by exponent, the Green-Kubo formula yields $\kappa=1016$ for $T=0.2$ and
$\kappa=2252$ for $T=200$. In order to get additional information another consuming simulation
is required. Still, from the other side, for $T=200$ at $N=640$ the logarithm of the heat
conductivity $\ln \kappa(N)=7.9>\ln (2252)=7.7$, and the dependence $\ln\kappa(N)$ (Fig.
\ref{fig4}, curve 2) does not demonstrate any trend towards convergence. Therefore the most
likely result in this case is also the divergence of the heat conductivity.

Let us consider the sequence $\kappa(N)$ ($N=10$, 20, 40, 80, 160, 320, 640) at other  values
of the cooperativeness. The results are summarized at Fig. \ref{fig7}.  The space of parameters
$(g,\tilde{T})$ is divided to two zones denoted by different colors. In the first (gray) zone
the sequence $\kappa(N)$ converges ($\kappa(160)\approx\kappa(320)\approx\kappa(640)$), and in
the second (white) zone the sequence grows monotonously.  Then, in the first zone
Frenkel-Kontorova model has finite heat conductivity, and in the second zone the heat
conductivity is either divergent or finite but very high (for $N\le 640$ the Fourier law is not
valid).

The first zone is limited by certain finite value of $g$: for some $g_0>1$ and for all  $g>g_0$
no convergence of $\kappa(N)$ was detected. The explanation is that for growing $g$ the system
becomes closer to continuum integrable sin-Gordon equation. At any fixed $g<g_0$ for $N\le 640$
the heat conductivity converges only for some finite temperature interval
$0<\tilde{T}_b<\tilde{T}<\tilde{T}_h<\infty$. As the cooperativeness decreases ($g\rightarrow
0$), the upper boundary of this interval tends to infinity ($\tilde{T}_h\rightarrow\infty$),
and the lower boundary tends to zero ($\tilde{T}_b\rightarrow 0$) proportionally to $g$.

\begin{figure}[t]
\begin{center}
\includegraphics[angle=0, width=0.85\textwidth]{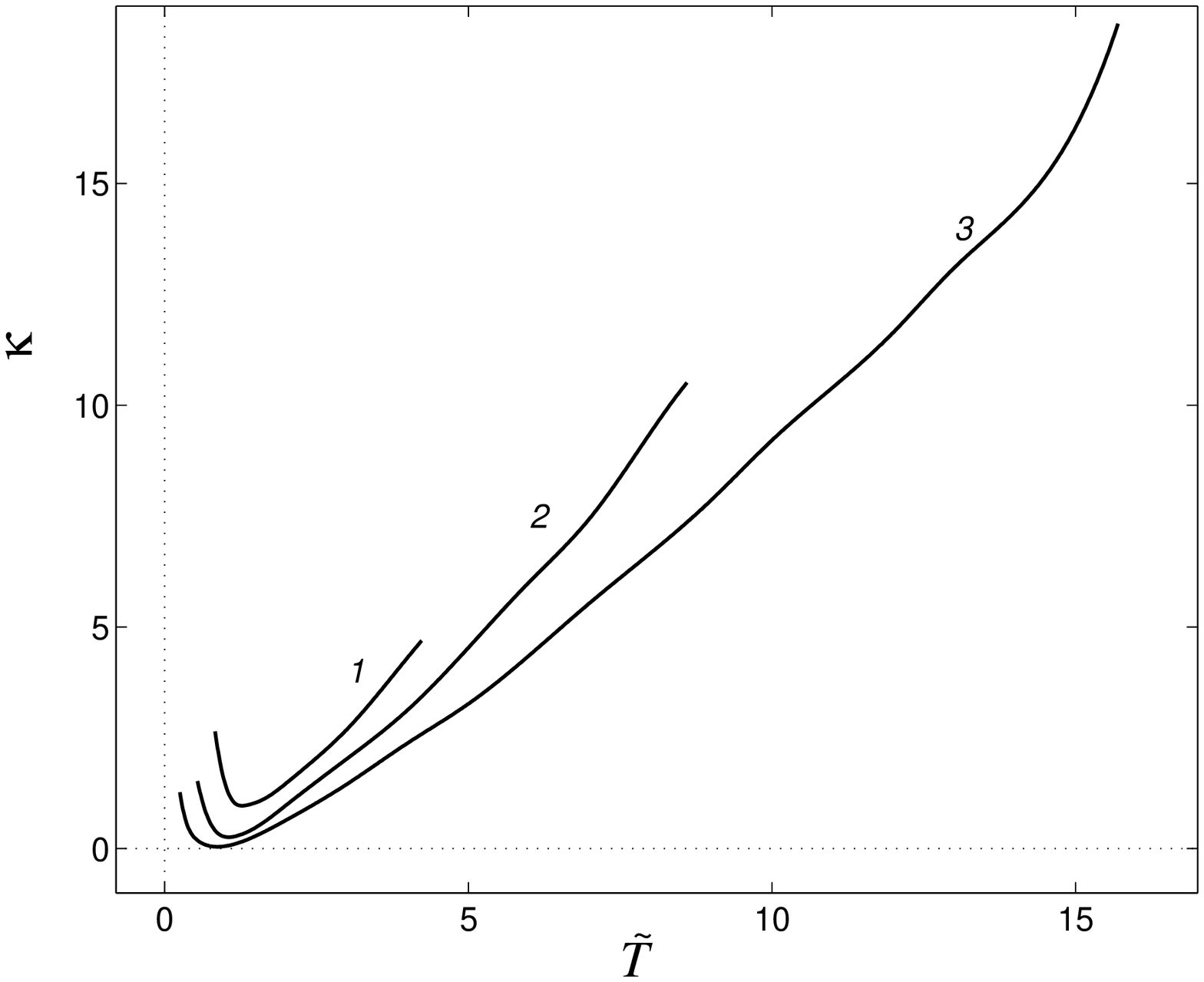}
\end{center}
\caption{\label{fig8}\protect \small
       Dependence of the heat conductivity coefficient $\kappa$ from
       the reduced temperature $\tilde{T}=T/\epsilon$ for the chain with
       periodic on-site potential (\ref{f4}) for
       $\epsilon=3$ (curve 1), $\epsilon=5$ (curve 2) and
       $\epsilon=10$ (curve 3).
        }
\end{figure}

The dependence of $\kappa$ on the reduced temperature $\tilde{T}$ is presented at Fig.
\ref{fig8}. Within the interval $[\tilde{T}_b,\tilde{T}_h]$ there exists a critical value
$\tilde{T}_m$ corresponding to the minimum of heat conductivity.

\begin{figure}[p]
\begin{center}
\includegraphics[angle=0, width=0.85\textwidth]{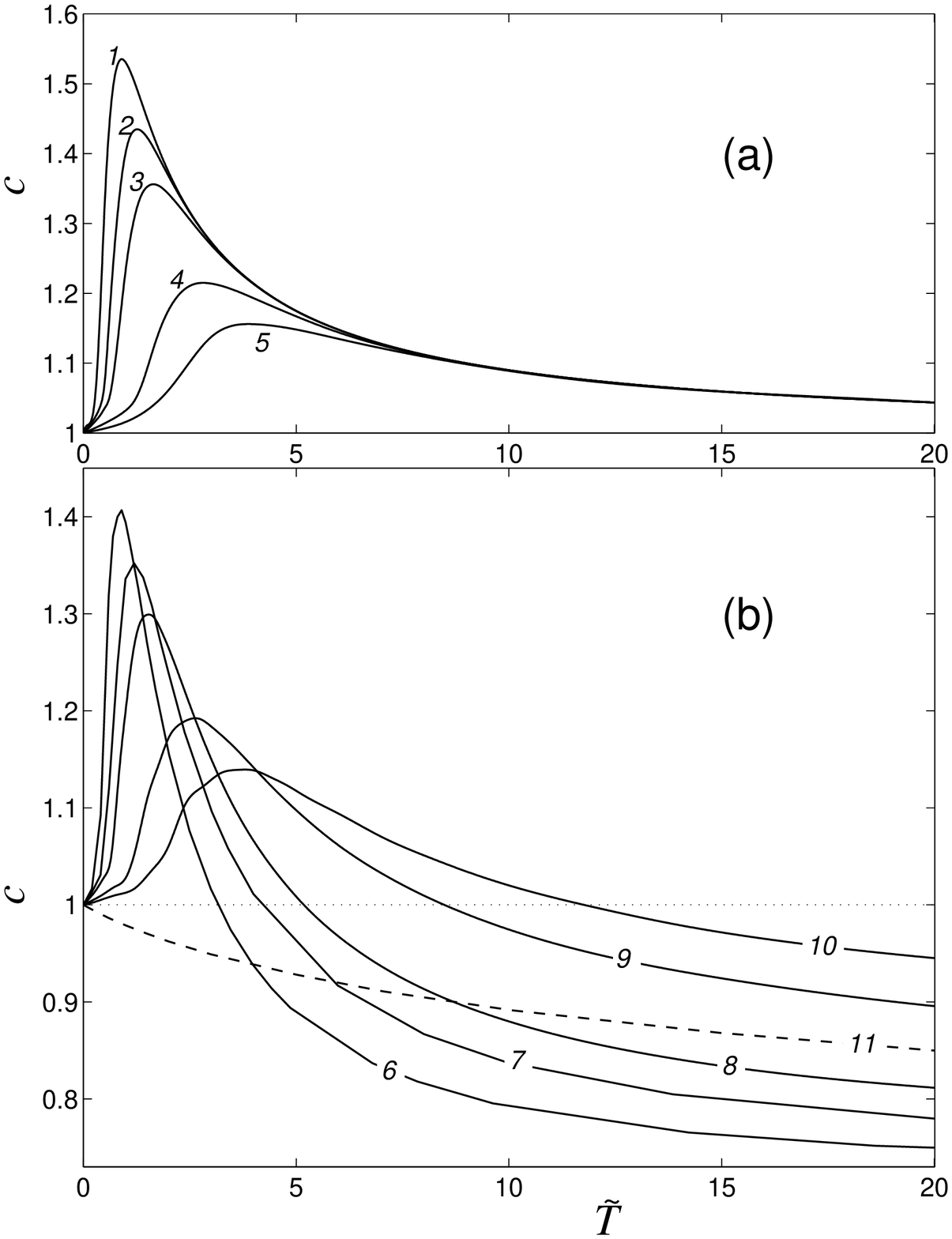}
\end{center}
\caption{\label{fig9}\protect \small
         The dependence of the dimensionless heat capacity $c$ on the reduced
         temperature $\tilde{T}=T/\epsilon$ (a) for the chain with periodic
         on-site potential (\ref{f4}) and (b) for the chain with $\phi$--4
         potential (\ref{f5}) for $\epsilon=10$ (curves 1,6), $\epsilon=5$ (curves 2, 7),
         $\epsilon=3$ (curves 3, 8), $\epsilon=1$ (curves 4, 9) and
         $\epsilon=0.5$ (curves 5, 10). Dashed curve 11 gives similar dependence
         for the chain with on-site sinh-Gordon potential (\ref{f6}), for $\omega_0=1$.
        }
\end{figure}

In order to reveal the mechanism of the heat conduction it is reasonable to explore the
behavior of heat capacity $c=\langle H\rangle/NT$ ($\langle H\rangle$ is the average energy of
cyclic $N$-atomic chain at the temperature $T$) on the reduced temperature $\tilde{T}$ (Fig.
\ref{fig9}). The heat capacity of classic harmonic chain is always unity, therefore the
discrepancy of this value from unity characterizes the significance of nonlinear effects at
given temperature. The lattice considered has negative anharmonism and therefore its heat
capacity must be more than unity for all temperature diapason. The heat capacity tends to unity
as $\tilde{T}\rightarrow 0$ and $\tilde{T}\rightarrow\infty$ and has single maximum at certain
temperature $\tilde{T}_c$. This value fairly coincides with the temperature $\tilde{T}_m$,
which corresponds to the minimum of the heat conductivity.

Moreover,  the increase and decrease of the heat capacity is clearly correlated with the
decrease and increase of the heat conductivity respectively. This fact allows suggesting the
same physical effects as responsible for both processes. For zero temperature the heat capacity
is equal to unity. The increase of the heat capacity at higher temperatures is related to
thermal activation of topological solitons (kinks and antikinks) which represent additional
degrees of freedom for this system. As a result the dynamical superlattice of solitons appears.
The density of this superlattice approaches its maximum at the temperature $\tilde{T}_m$.
Further growth of the temperature results in the decrease of the number of degrees of freedom,
which is manifested as effective detaching of the chain from the on-site potential. Therefore
the heat capacity decreases and tends to unity as the temperature grows.

\begin{figure}[t]
\begin{center}
\includegraphics[angle=0, width=0.85\textwidth]{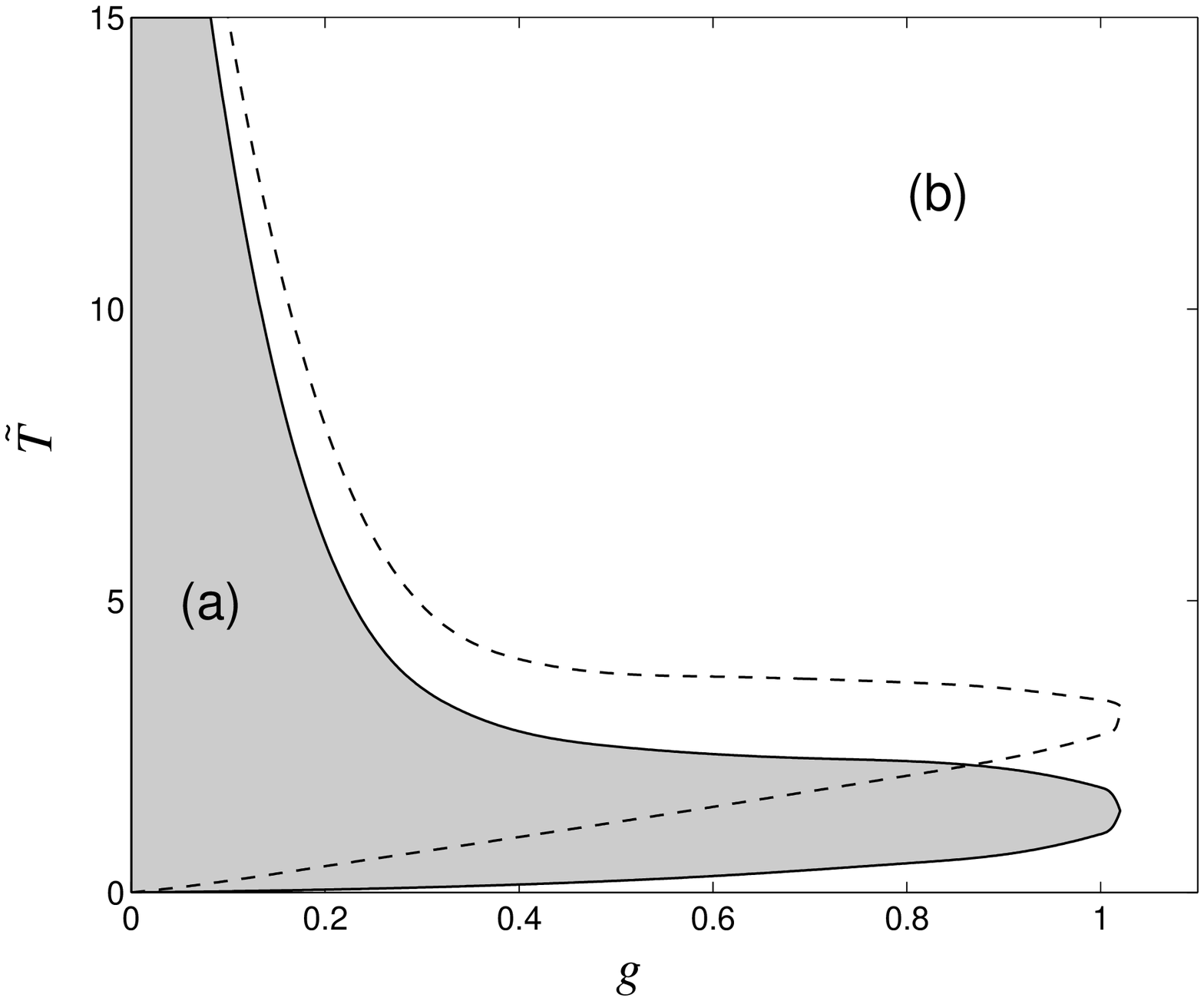}
\end{center}
\caption{\label{fig10}\protect \small
       Zones in the space of parameters $(g,\tilde{T})$, where for
       finite chains $N\le 640$ with periodic on-site potential
       (\ref{f4}) and period $l=2\pi\sqrt{2}$
       the heat conductivity is normal (a, gray) and abnormal (b, white).
        Dashed line denotes the same boundary for commensurate Frenkel-Kontorova model ($l=2\pi$).
        }
\end{figure}

Correlations between the behavior of the heat capacity and the heat conductivity and especially
fair coincidence of $\tilde{T}_m$ and $\tilde{T}_c$ allow us to suppose that the heat transfer
is limited by phonon scattering on the soliton superlattice. The effectiveness of such
scattering depends on the density of the superlattice as well as on the ability of single kink
to scatter phonons. In the strongly cooperative regime $g>g_0$ the interaction between solitons
and phonons is nearly elastic (close to the case of complete integrability) and therefore the
heat conductivity has the trend to diverge. For lower cooperativeness the soliton-phonon
interaction is less elastic and finite temperature diapason $[\tilde{T}_b,\tilde{T}_h]$ of
converging heat conductivity appears. For the cases of low $\tilde{T}<\tilde{T}_b$ and high
$\tilde{T}>\tilde{T}_h$ temperatures the convergence of the heat conductivity cannot be detected
in the framework of current experiment. The suggested reason of this effect is that the soliton
superlattice effectively disappears.

Let us consider  now incommensurate Frenkel - Kontorova chain where the period of the chain is
different from the period of on-site potential. The dimensionless on-site potential is periodic
function(\ref{f4}) with period $2\pi$, and the chain has period $l=2\pi q$. Then in system of
equations (\ref{f7}) function $F(u_n)$ will take the form
$$
F(u_n)=\frac{d}{du} U(u_n+nl).
$$
For the sake of simulation we choose $q=l/2\pi=\sqrt{2}$, corresponding  in certain sense to
extremely incommensurate case. It is well-known \cite{p20} that such a lattice in its ground
state already has soliton superlattice of nonzero density. Therefore the convergence of the
heat conductivity is expected to be facilitated as compared to the commensurate case.

Fig. \ref{f10} demonstrates the zone in the space of parameters $(g,\tilde{T})$ where  the
sequence $\kappa(N)$, $N=10$, 20, 40, 80, 160, 320, 640. converges. For the sake of comparison
the boundary for the commensurate case is also presented ($l=2\pi$). The result is than no
qualitative change of the behavior occurs. The only difference is that the zone with normal
heat conductivity moves downwise. This effect is related to presence of superlattice of
solitons at any temperature. The transition to normal heat conduction occurs at lower
temperature since less solitons should be thermally activated in order to achieve convergence.
From the other side, the soliton superlattice facilitates effective detachment of the lattice
from the on-site potential (the average coupling energy in the ground state is less) and
therefore the upper boundary for the normal heat conduction is also achieved at lower
temperatures.

\section{Heat conductivity of the chain with double-well on-site potential}

Let us consider the heat conductivity of the chain with on-site potential $\phi$-4  (\ref{f5}).
For this case the analysis of the sequence $\kappa(N)$, $N=10$, 20, 40, 80, 160, 320, 640,
demonstrates that the heat conductivity converges as $\tilde{T}>\tilde{T}_0=3g/2$ ($T>1.5$) -
see Fig. \ref{fig7}.
\begin{figure}[t]
\begin{center}
\includegraphics[angle=0, width=0.85\textwidth]{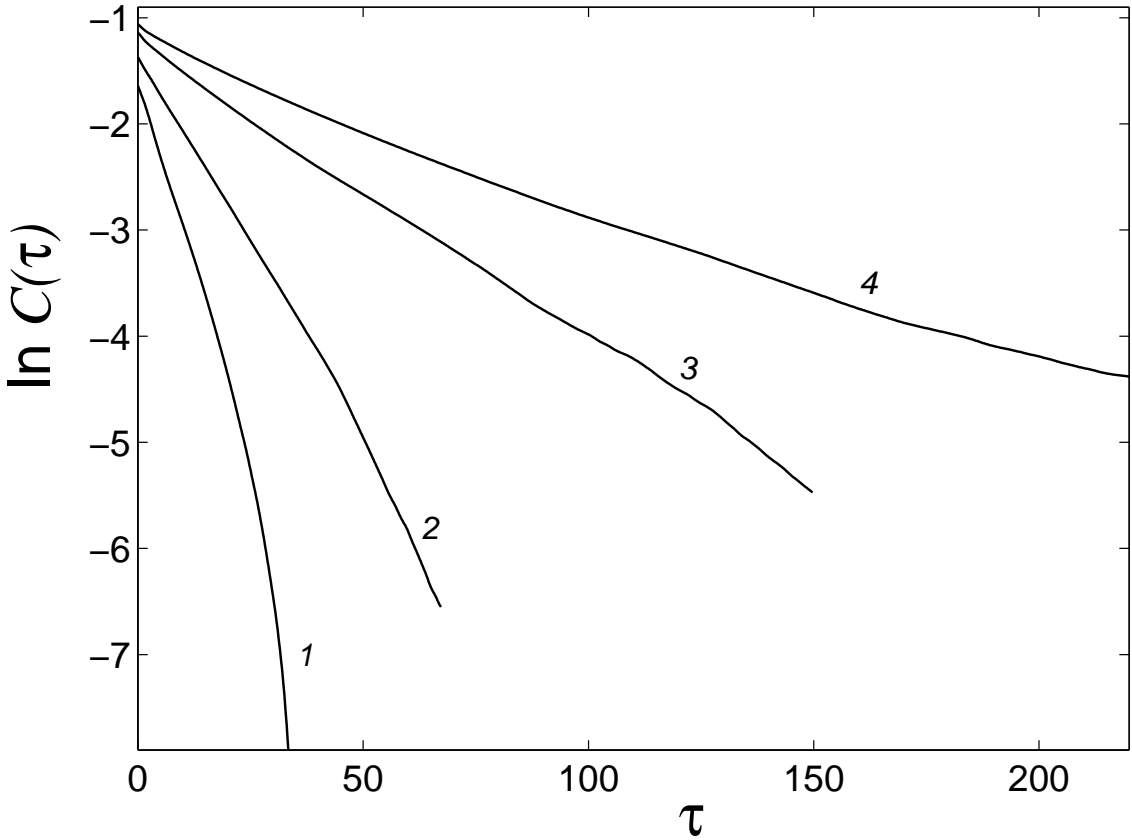}
\end{center}
\caption{\label{fig11}\protect \small
       Exponential decrease of the autocorrelation function $C(\tau)$
       in the chain with on-site potential $\phi$--4 (\ref{f5}), $\epsilon=1$,
       $T=20$ (curve 1), $T=10$ (curve 2), $T=5$ (curve 3) and
       $T=3$ (curve 4).
       (Semilogarythmic coordinates $\ln C(\tau)$ versus $\tau$).
        }
\end{figure}

\begin{figure}[t]
\begin{center}
\includegraphics[angle=0, width=0.85\textwidth]{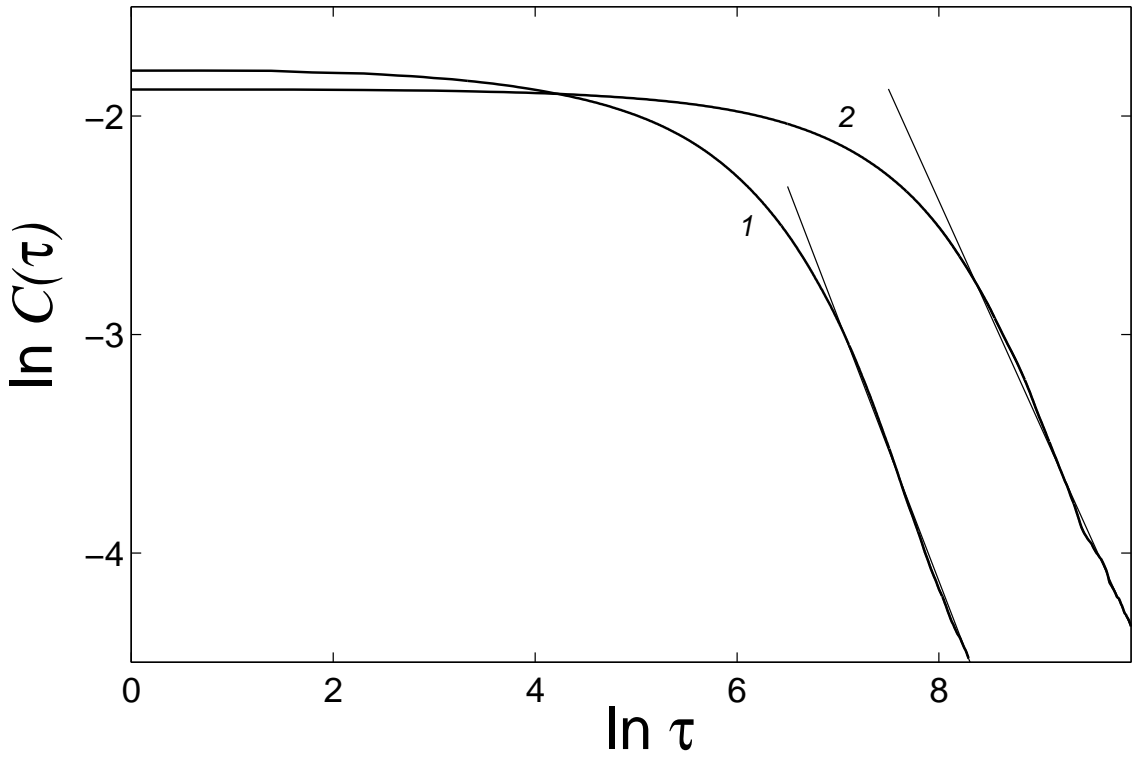}
\end{center}
\caption{\label{fig12}\protect \small
       Power-law decrease of the autocorrelation function $C(\tau)$ in the chain with
       $\phi$--4 on-site potential (\ref{f5}), $\epsilon=1$, $T=1$ (curve 1) and $T=0.5$
       (curve 2). (Double logarithmic coordinates, $\ln C(\tau)$ versus $\ln \tau$).
       The angle coefficient $\alpha$ determines the decrease rate. For $T=1$
       $\alpha=1.2$, for $T=0.5$ $\alpha=1.02$.
        }
\end{figure}

In order to investigate the character of the heat conduction in the temperature range
$\tilde{T}<\tilde{T}_0$ let us consider the temperature behavior of the autocorrelation
function $C(\tau)$. For $g=1$ ($\epsilon=1$) this behavior is demonstrated at Fig. \ref{fig11}.
As $\tau\rightarrow\infty$ the autocorrelation function decreases exponentially. The decrease
rate grows as the temperature increases and therefore the conclusion concerning finite heat
conductivity at $\tilde{T}>\tilde{T}_0$ is confirmed. At lower temperatures the decrease rate
satisfies a power law $\tau^{-\alpha}$ -- see Fig. \ref{fig12}.  The degree $\alpha$ decreases
with the decrease of the temperature. At $T=1$ $\alpha=1.2>1$, therefore integral (\ref{f15})
converges and the heat conductivity is finite, at $T=0.5$ $\alpha=1.02$. Within the accuracy
available for current numerical possibilities this value corresponds to transition to abnormal
heat conduction. It is extremely difficult to obtain reliable  data for lower temperatures in
order to substantiable this conclusion because of huge computation time required. The reason is
that the system is rather close to completely integrable case. New numerical methods based on
the latter fact are desirable for investigation of this kind of systems.

\begin{figure}[t]
\begin{center}
\includegraphics[angle=0, width=0.85\textwidth]{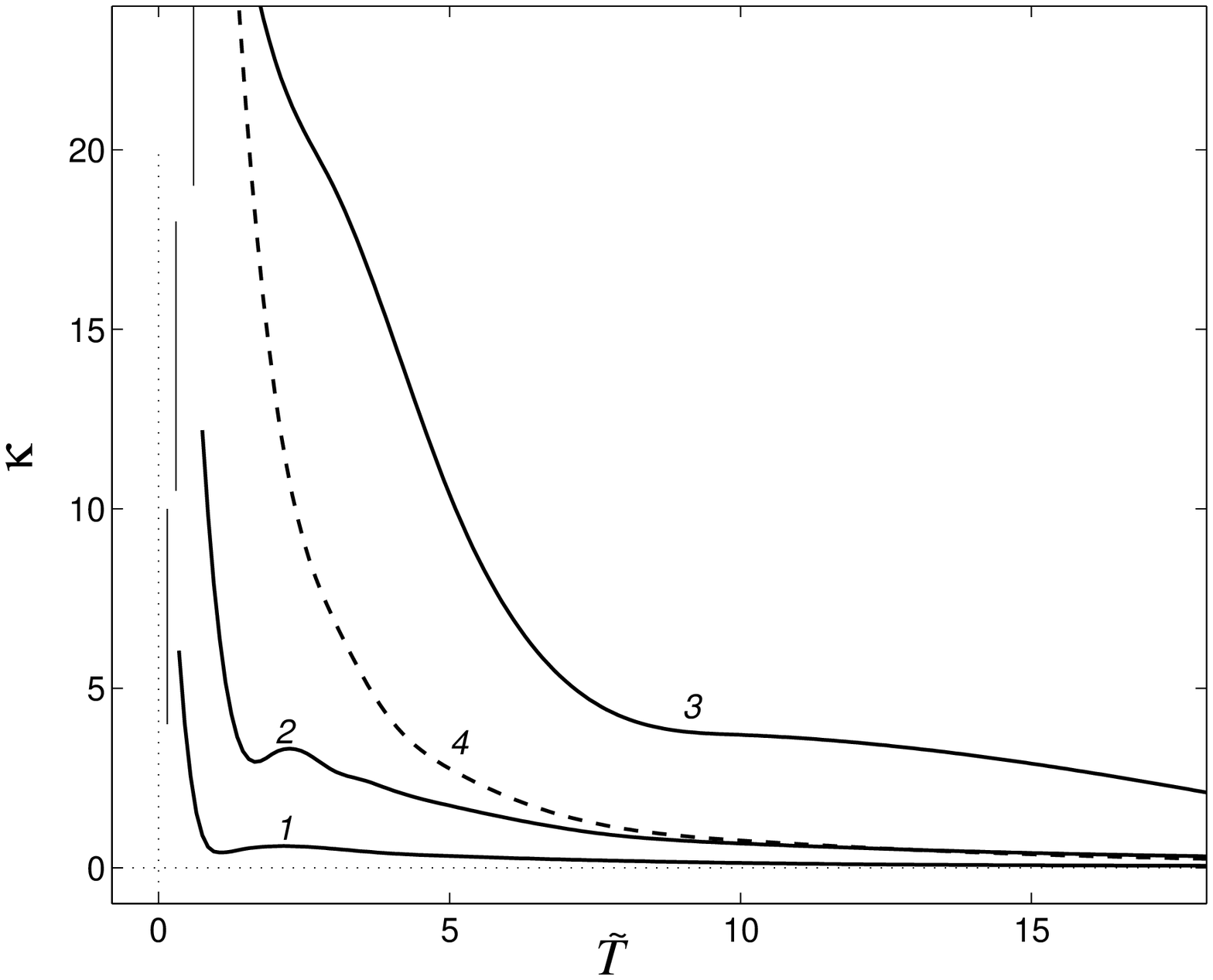}
\end{center}
\caption{\label{fig13}\protect \small
       Heat conductivity $\kappa$ versus reduced temperature $\tilde{T}=T/\epsilon$
       for the chain with double-well on-site potential (\ref{f5}), $\epsilon=4$ (curve 1),
       $\epsilon=2$ (curve 2) and $\epsilon=1$ (curve 3) and for the chain with
       sinh-Gordon on-site potential (\ref{f6}), $\omega_0=1$ (dashed curve 4, $\tilde{T}=T$).
        }
\end{figure}

The dependence of the heat conductivity $\kappa$ on reduced temperature $\tilde{T}$ is
presented at Fig. \ref{fig13}. For low cooperativeness ($g<0.5$) the heat conductivity
approaches local minimum and afterwards local maximum at $(\tilde{T}=\tilde{T}_1)$ and
monotonously decreases to zero as $\tilde{T}\rightarrow\infty$. The relative value of the
maximum decreases as the cooperativeness grows and disappears for certain critical value of
$\epsilon$.

In  order to reveal the physical reasons of such behavior of the heat conductivity it is also
reasonable to investigate the behavior of the heat capacity $c$ (Fig. \ref{fig9}b). As
$\tilde{T}\rightarrow 0$ the heat capacity $c\rightarrow 1$. As the temperature grows, the heat
capacity grows, achieves its maximum at the temperature $\tilde{T}_c$ and then decreases
monotonously to the value less than unity. The value $\tilde{T}_c$ is situated near the maximum
point of the heat conductivity $\tilde{T}_1$. Such behavior is related with the peculiarities
of $\phi$--4 potential. At low temperatures the main effect is due to negative anharmonism near
the ground state (therefore the heat capacity exceeds unity) and for high temperatures
($\tilde{T}\gg 1$) the process is governed by positive anharmonism bringing the heat capacity
to the value below unity.

\begin{figure}[p]
\begin{center}
\includegraphics[angle=0, width=0.85\textwidth]{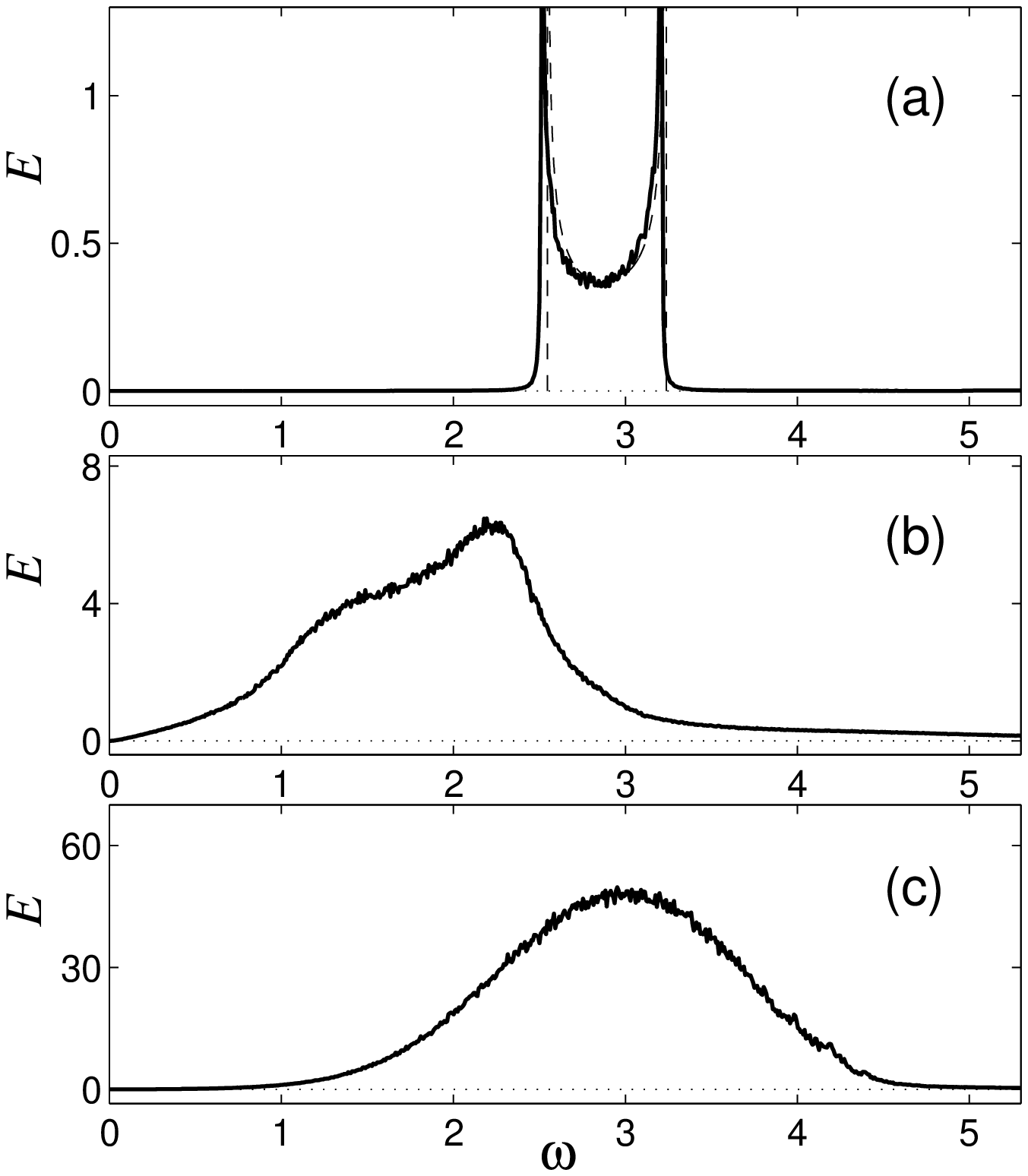}
\end{center}
\caption{\label{fig14}\protect \small
       Frequency spectrum of energy of vibrations in the chain with double-well
       on-site potential (\ref{f5}) at temperatures $T=0.4$ (a), $T=10$ (b) and $T=100$ (c).
       $\epsilon=4$. Dashed line denotes the spectrum of harmonic chain (\ref{f3})
       with $\omega_0=4/\pi\sqrt{\epsilon}$.
        }
\end{figure}

Let us now consider the frequency spectrum of vibrations of the chain. The spectrum is computed
for $\epsilon=4$ $(g=1/4)$ and three characteristic temperatures $T=0.4$, 10, 100. The spectrum
of the chain with harmonic on-site potential (\ref{f3}) does not depend on the temperature and
has the form
       \begin{equation}
       E(\omega)=2\omega/\pi \sqrt{(\omega^2-\omega_0^2)(\omega_1^2-\omega^2)},
       \label{f17}
       \end{equation}
where  maximum frequency $\omega_1^2=4+\omega_0^2$. For $\epsilon=4$,
$\omega_0=4/\pi\sqrt{\epsilon}=2.546$, $\omega_1=3.238$. As it is demonstrated at Fig.
\ref{fig14}a, for temperature $T=0.4$ the spectrum of the chain with on-site $\phi$--4
potential nearly coincides with the spectrum of purely harmonic chain (\ref{f17}). Such
spectrum means that at low temperatures only thermalized phonons contribute to the frequency
spectrum and other excitations do not play any significant role. For $T=10>\epsilon$ the
distribution crosses the lower boundary of the propagation zone $\omega_0$ (Fig. \ref{fig14}b).
Such low-frequency component may be associated with intrinsic vibrations of the solitons
superlattice. For even higher temperatures $T=100\gg\epsilon$ the spectrum crosses also the
upper boundary of the propagation zone $\omega_1$ (Fig. \ref{fig14}c). Such effect may be
attributed only to thermalization of high-frequency discrete breathers.
Therefore,  for low temperatures $\tilde{T}<\tilde{T}_0=0.5g$ the dynamics of the system is
close to that of harmonic chain. The heat transport is governed by weakly interacting phonons
and heat conductivity diverges. For higher temperatures the heat conductivity converges. In the
intermediate diapason $\tilde{T}_0<\tilde{T}<\tilde{T}_1$ the effective phonon scattering
mechanism exists due to the superlattice of topological solitons, and for high temperatures
$\tilde{T}>\tilde{T}_1$ -- due to high - frequency discrete breathers. Interplay of two
different mechanisms of the phonon scattering explains also the dependence of the heat
conductivity on the cooperativeness of the system (Fig. \ref{fig13}). The minimum and maximum
of heat conductivity disappear with growth of the cooperativeness since the soliton mechanism
of scattering becomes less effective (the soliton-phonon interaction is closer to elastic) and
simultaneously the excitation of the discrete breathers becomes easier.

\section{Heat conductivity of the chain with sinh-Gordon on-site\\ potential}

\begin{figure}[p]
\begin{center}
\includegraphics[angle=0, width=0.75\textwidth]{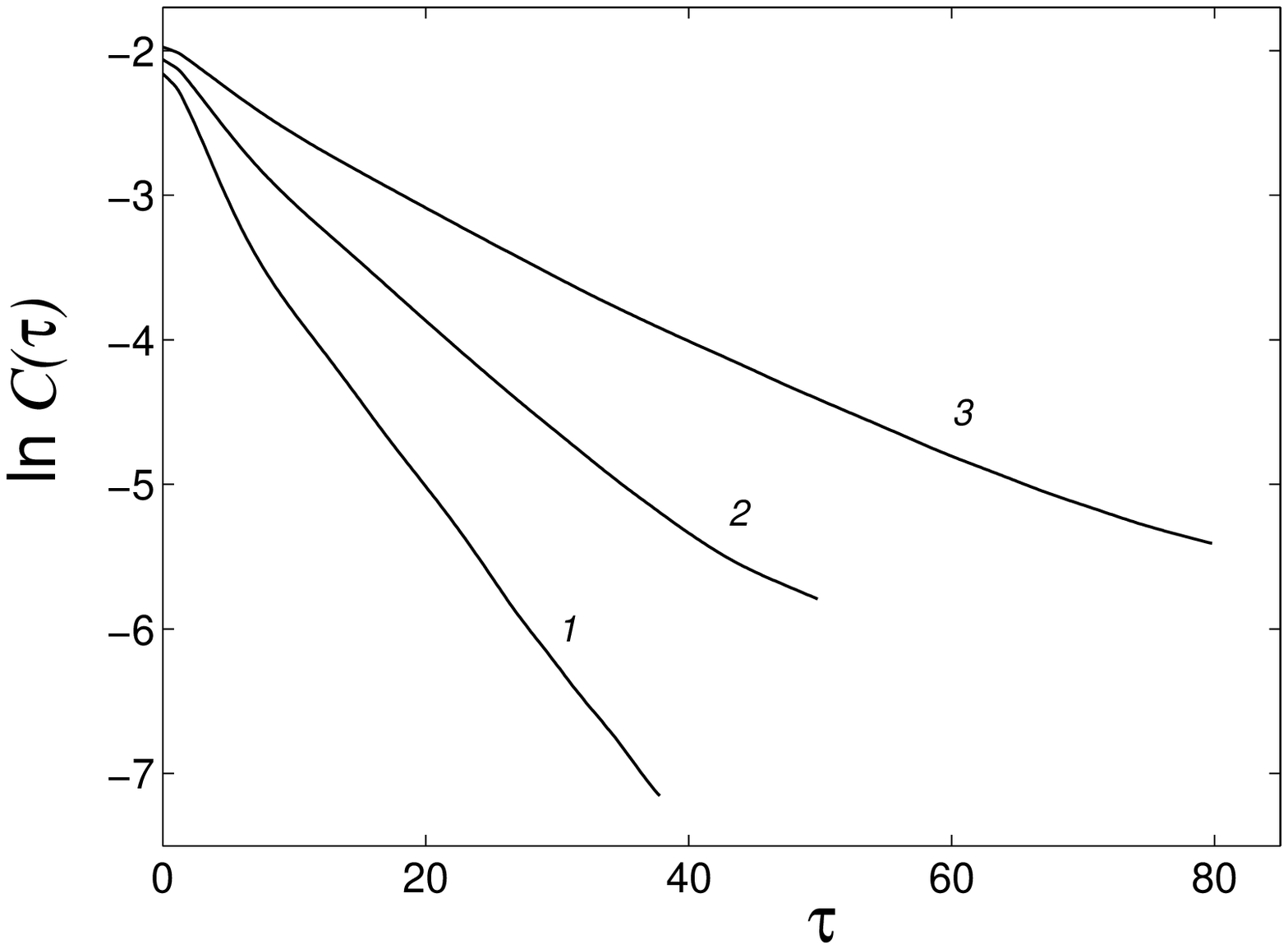}
\end{center}
\caption{\label{fig15}\protect \small
       Exponential decrease of the autocorrelation function $C(\tau)$
       in the chain with sinh-Gordon on-site potential (\ref{f6}), $\omega_0=1$,
        $T=10$ (curve 1), $T=7$ (curve 2) and $T=5$ (curve 3).
       (Semilogarythmic coordinates $\ln C(\tau)$ versus $\tau$).
        }
\end{figure}

\begin{figure}[p]
\begin{center}
\includegraphics[angle=0, width=0.75\textwidth]{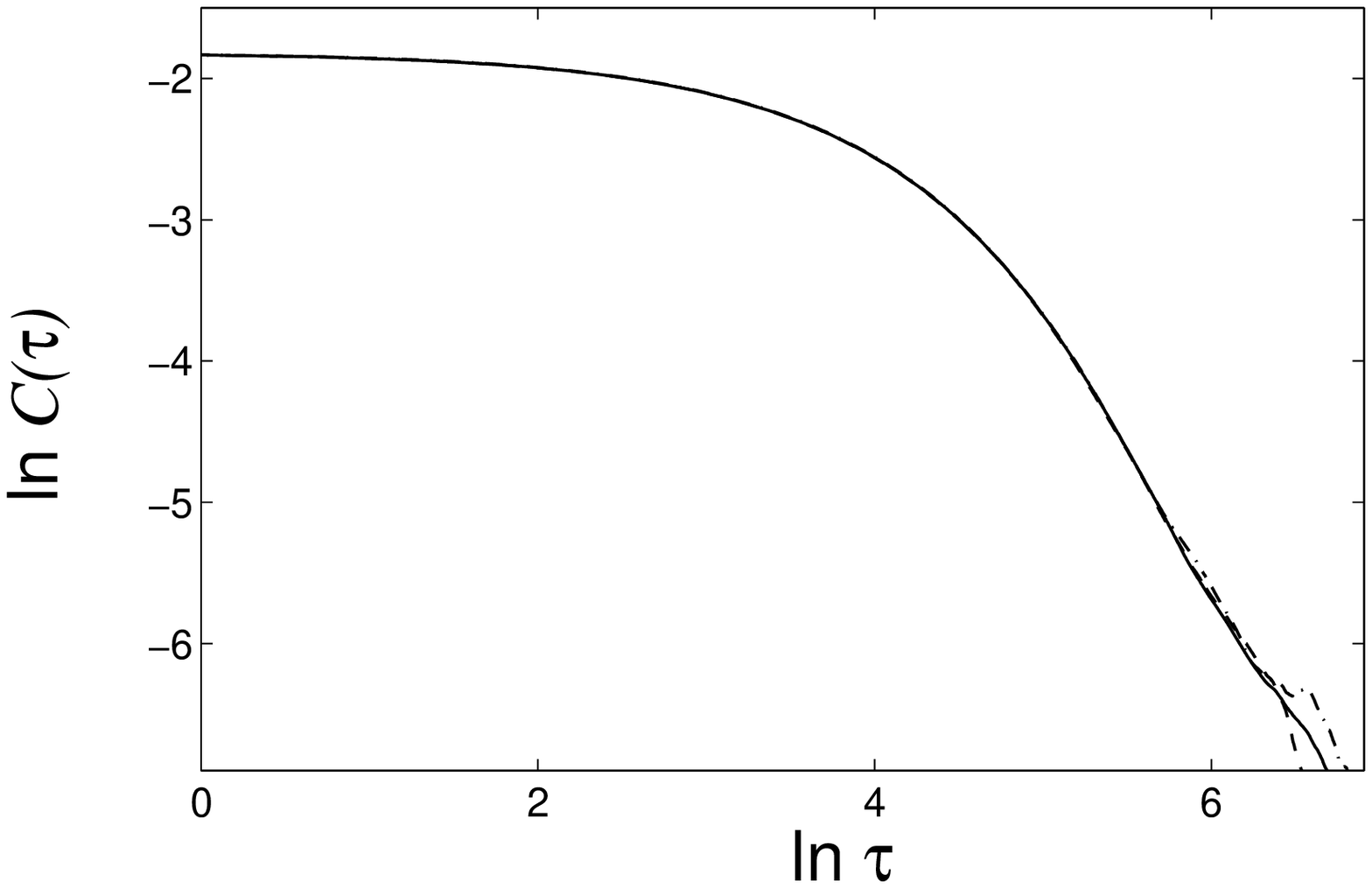}
\end{center}
\caption{\label{fig16}\protect \small
       Power-law decrease of the autocorrelation function $C(\tau)$ in the chain
       with sinh-Gordon on-site potential (\ref{f6}), $\omega_0=1$, $T=2$. Solid
       line corresponds to the number of particles $N=500$, dotted -- to $N=1000$,
       dashed-dotted -- to $N=2000$.
        }
\end{figure}

Heat conductivity of this system has been investigated in paper \cite{p10} and here it is
reasonable to elucidate the details related to physical mechanisms of the process. The on-site
potential (\ref{f6}) is single-well function with positive anharmonism. The analysis of the
finite sequence $\kappa(N)$, $N=10$, 20, 40, 80 160, 320, 640, demonstrates that the heat
conductivity converges for high temperatures ($T>T_0>0$).  This observation is supported by the
fact that the autocorrelation function $C(\tau)$ at high temperatures for
$\tau\rightarrow\infty$ decreases exponentially (Fig. \ref{fig15}), and for low temperatures -
by power law (Fig. \ref{fig16}).

The heat conductivity decreases monotonously and for $T\rightarrow\infty$ exponentially tends
to zero (Fig. \ref{f13}, curve 4). Positive anharmonism of the potential leads to monotonous
decrease of the heat capacity (Fig. \ref{f9}, curve 11). The frequency spectrum of vibrations
moves towards the upper boundary of the propagation zone with growth of the temperature. These
facts allow concluding that the high-frequency discrete breathers provide effective phonon
scattering in this model and facilitate the convergence of the heat conductivity. Growing
concentration of these breathers with the growth of the temperature leads to monotonous
decrease of the heat conductivity coefficient.

Chain with on-site potential
       \begin{equation}
       V(u)=\beta u^4/4 \label{f18}
       \end{equation}
( positive $\phi^4$ model) also has finite heat  conductivity \cite{p10a,p10b}. Potential
(\ref{f18}) as well as sinh-Gordon on - site potential (\ref{f6}) is single - well symmetric
function with positive anharmonism. Therefore the mechanism of the phonon scattering is also
related to the discrete breathers and $\kappa(T)\searrow 0$ for $T\rightarrow\infty$. For
$\beta=2$ the heat conductivity $\kappa(T)\sim T^{-1.35}$ \cite{p10b}.

\section{Conclusion}

The investigation presented above demonstrates  that the anharmonicity of the on-site potential
does not constitute sufficient condition for the convergence of the heat conductivity
coefficient. The behavior of any concrete model in the above respect depends on its peculiar
nonlinear excitations which determine the process of the heat transfer and phonon scattering.
Two typical mechanisms of the phonon scattering were revealed in the paper  -- thermalized
soliton superlattice (discrete sin-Gordon  and $\phi$--4 models) and discrete high-frequency
breathers ($\phi$--4 and sinh-Gordon models). Phonon scattering mechanism may switch with the
change of the temperature ($\phi$--4 -- model).

For the discrete Frenkel-Kontorova model the zone of the converging heat conductivity for given
chain length is limited by low and high temperatures and by high cooperativeness. The numerical
possibilities available up to date does not allow establishing unambiguously the character of
the heat conductivity outside the zone designated at Fig. \ref{fig7}. Still there is a reason
to suggest that infinite chain for certain parameters has diverging heat conductivity, although
the zone corresponding to finite heat conductivity will be larger that computed above.

Unlike Frenkel-Kontorova model, for $\phi$--4 model it is possible to demonstrate that for low
temperatures the boundary of the transition to abnormal heat conduction may be achieved. It is
possible to suppose that there exist a transition from infinite to finite heat conductivity
with growth of temperature for any cooperativeness.  The probable reason for the divergence to
be detectable is the presence  of odd-power terms in the on-site potential in the vicinity of
the extrema of the potential wells.

The sinh-Gordon model does not allow to detect the divergence of the heat conductivity in
current experiment; still, the transition also may be suggested for any cooperativeness.

It is possible to suggest that for any analytic on-site potential for low temperatures the heat
conductivity will diverge.

The authors are grateful to Russian foundation of Basic Research (grant 01-03-33122),
to RAS Commission for Support of Young Scientists (6th competition, grant no. 123) and
to Fund for Support of Young Scientists for financial support.

A.V. Savin is grateful to International Association of Assistance for the promotion of
co-operation with scientists from the New Independent States of the former Soviet Union
(project INTAS no. 96-158) for financial support.

\section{Appendix}

\subsection{Numeric realization of the Langevin thermostat \label{pr1}}

System of equations describing the dynamics of the chain attached to thermostats (\ref{f7}) has
been integrated numerically by standard fouth-order Runge-Kutta method with constant step of
integration $\Delta\tau$. Numeric realization of delta-function is performed as
$\delta(\tau)=0$ for $|\tau|>\Delta\tau/2$ and $\delta(\tau)=1/\Delta\tau$ for
$|\tau|\le\Delta\tau$, i.e. the step of integration corresponds to the correlation time of the
random forces. That is why in order to get correct description of the Langevin thermostat we
must guarantee that the relaxation time $\tau_r\gg \Delta\tau$. In order to fulfill this
condition the relaxation time was chosen as $\tau_r=10$, and the step of integration for
different values of $N$, was chosen as $\Delta=0.05$, 0.025, 0.0125.

For every step of integration the random forces $\xi_n^\pm$ were taken to be constant. They
were computed as independent realizations of the random value $\xi$, normally distributed with
zero average $\langle\xi\rangle=0$ and dispersion
$\langle\xi^2\rangle=2T_\pm/\tau_r\Delta\tau$. For generating the random value $\xi$ program
package ZUFALL \cite{p21} was used.

The initial state for the integration of equations (\ref{f7}) was chosen to be equal to ground
state of the chain:
           \begin{equation}
           \label{f19}
           u_n=u_0,~~{u_n}'=0,~~n=1,2,....,N_++N+N_-,
           \end{equation}
where $u_0=0$  for on-site potentials (\ref{f3}), (\ref{f6}) and $u_0=\pi$ for potentials
(\ref{f4}), (\ref{f5}). It is convenient to control the accuracy of the simulation through
behavior of sequence of average local heat fluxes $\{J_n\}_{n=N_++1}^{N_++N}$. If the choice of
the integration step $\Delta\tau$ is correct then this sequence should be constant. If the
local average heat flux changes from particle to particle then the integration step should be
reduced. For growing chain length $N$ the step of integration should be also reduced in order
to provide sufficient accuracy; the averaging time also grows (see \cite{p22}) and therefore
the time of simulation necessary for obtaining reliable results for large $N$ turns out to be
extremely large.

\subsection{Computation of the correlation function \label{pr2}}

In order  to compute the autocorrelation function of the heat flux $C_N(\tau)$ dynamics of
cyclic $N$-particle chain was simulated. The thermalized chain with temperature $T$ was
obtained by integrating Langevin system of equations
        \begin{eqnarray}
        u''_n&=&u_{n+1}-2u_n+u_{n-1}-F(u_n)-\gamma u'_n+\xi_n, \label{f20} \\
           && n=1,2,...,N~,  \nonumber
        \end{eqnarray}
where  $n+1=1$ for $n=N$ and $n-1=N$ for $n=1$, $\gamma=0.1$ (relaxation time $\tau_r=10$),
$\xi_n$ -- white Gaussian noise normalized as
        $$
        \langle\xi_n(\tau)\rangle=0,~~~
        \langle\xi_n(\tau_1)\xi_k(\tau_2)\rangle
        =2\gamma T\delta_{nk}\delta(\tau_2-\tau_1).
        $$
System (\ref{f20}) has been integrated numerically with initial conditions corresponding to the
ground state of the chain. After time $\tau=10\tau_r$ the chain approached equilibrium with the
thermostat and the coordinates
        \begin{equation}
        \{u_n(\tau),{u_n}'(\tau)\}_{n=1}^N \label{f21}
        \end{equation}
corresponding to the thermalized state at temperature $T$.

Afterwards  the dynamics of isolated thermalized chain was simulated. For this sake system
(\ref{f20}) was integrated with zero damping $\gamma=0$ and zero external force $\xi_n\equiv
0$). Thermalized state (\ref{f21}) was used as initial condition. The result was the dependence
of the general heat flux {\bf J} on time $\tau$. Afterwards with the help of (\ref{f16}) the
autocorrelation function $C_N(\tau)$ was computed for given thermalized state of the chain. The
autocorrelation function depends significantly on concrete realization of the thermalized
chain. That is why in order to improve the accuracy this procedure was performed $10^3\div
10^4$ with independent initial realizations of the thermalized state. Finally the shape of the
correlation function was computed as average over all these realizations. It is worth while
mentioning that the alternative way of computation (performing of one very large simulation)
would not bring about any sufficient gain in the accuracy because of growing integration
errors.

In order to verify the independence of the correlation function on the chain length the
appropriate calculations were performed for different values of $N$. Fig. \ref{fig16}
demonstrates the function $C_N(\tau)$ for the chain with sinh-Gordon on-site potential for
$\omega_0=1$, $T=2$ and $N=500$, 1000, 2000. It is clear that the autocorrelation function is
nearly independent on $N$ (the differences are noticeable only for large times and are reduced
as the number of realizations used for averaging grows).  For given set of parameters $N=1000$
provides sufficient accuracy.

\subsection{Comparison of Langevin and Nose-Hoover thermostats \label{pr3}}

Unlikely the Langevin thermostat, the Nose-Hoover thermostat (NHT)\cite{p16} is not stochastic.
Its dynamics is completely determined by the initial conditions. It turned out to be good
choice for simulations of FPU system \cite{p2,p3} but its deterministic nature can bring about
artifacts in the behavior of the system. We compare this thermostat with the Langevin
thermostat (LT) we use for the case of Frenkel - Kontorova model.

Let us consider the chain with fixed ends ($1<n<N+N_++N_-$) with $N_\pm$ particles attached to
NHT having the temperature $T_\pm$. Dynamics of the system is described by equations
        \begin{eqnarray}
        u''_n&=&u_{n+1}-2u_n+u_{n-1}-F(u_n)-\eta_+u'_n,
        \nonumber \\
        n&=&2,...,N_+,\nonumber \\
        {\eta_+}'&=&\frac{1}{\tau_r^2}
        \left(\frac{1}{(N_+-1)T_+}\sum_{n=2}^{N_+}{u_n'}^2-1\right)
        \nonumber\\
        u''_n&=&u_{n+1}-2u_n+u_{n-1}-F(u_n),
        \label{f22} \\
        n&=&N_++1,...,N_++N, \nonumber \\
        u''_n&=&u_{n+1}-2u_n+u_{n-1}-F(u_n)-\eta_-u'_n,
        \nonumber \\
        n&=&N_++N+1,...,N_++N+N_--1,\nonumber \\
        {\eta_-}'&=&\frac{1}{\tau_r^2}
        \left(\frac{1}{(N_--1)T_-}\sum_{n=N_++N+1}^{N_++N+N_-}{u_n'}^2-1\right)
        \nonumber
        \end{eqnarray}
where $F(u)=dU(u)/du$, and $\tau_r$ is the relaxation time of the thermostat.

\begin{figure}[p]
\begin{center}
\includegraphics[angle=0, width=0.85\textwidth]{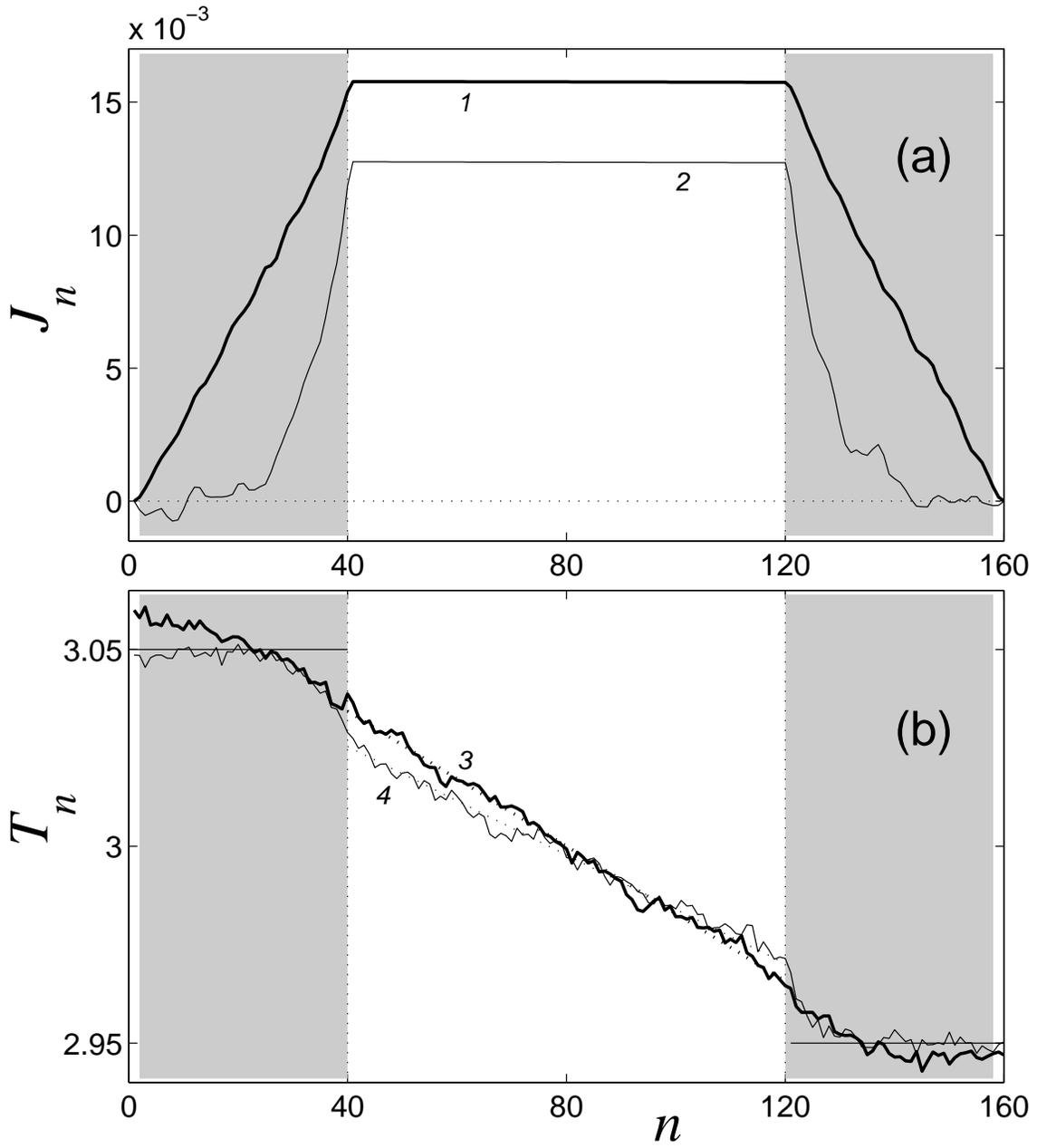}
\end{center}

\caption{\label{fig17}\protect \small
       Distribution of local heat flux $J_n$ (a) and
       and local temperature $T_n$ (b) in the chain with periodic on-site potential (\ref{f4}),
       $\epsilon=1$, $N=160$, $N_\pm=40$, $T_+=3.05$, $T_-=2.95$,
       averaging time $\tau=10^7$. Grey zones denote the chain fragments embedded in the
       thermostats. Thick lines (1, 3) correspond to NHT ($\tau_r=1$), and thin (2, 4)
       -- to LT ($\tau_r=10$).
        }
\end{figure}
\begin{figure}[tbh]
\begin{center}
\includegraphics[angle=0, width=0.85\textwidth]{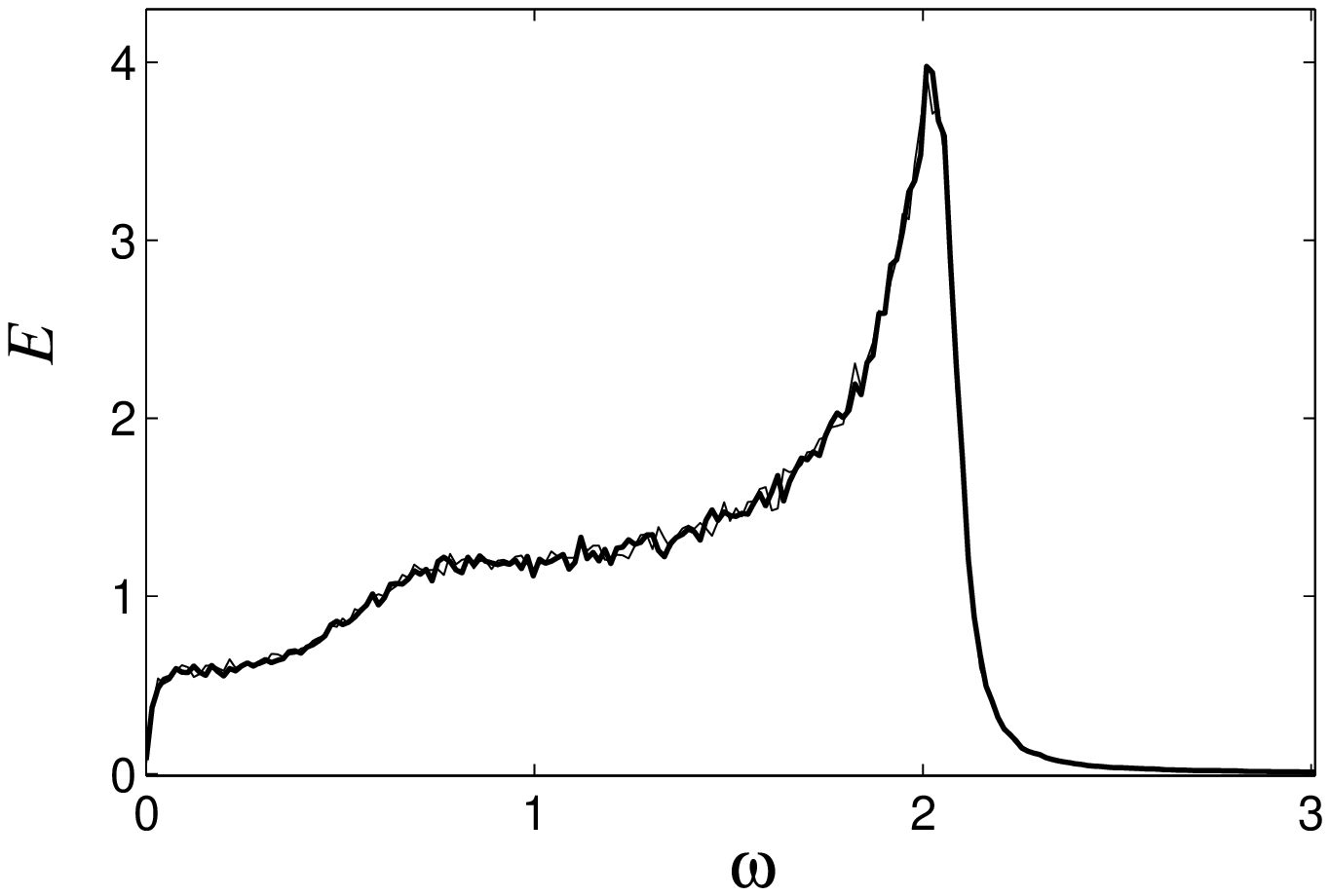}
\end{center}
\caption{\label{fig18}\protect \small
       Frequency distribution of the energy of particle having number $N/2$ in the chain
       with periodic on-site potential (\ref{f4}), $\epsilon=1$, $N=160$, $N_\pm=40$,
       $T_+=3.05$, $T_-=2.95$. Thick line corresponds to use of NHT ($\tau_r=1$), and thin
       -- to use of LT ($\tau_r=10$).
        }
\end{figure}

Usually the simulations of the heat conductivity \cite{p2,p3,p9,p10a} take $\tau_r=1$, and
$N_+=N_-=2$ (only end particles are attached to the thermostat $n=2$ and $n=N_++N+N_--1$). But,
as stated in \cite{p17}, such thermostats are not enough random -- they cover only a part of
the phase space and correspond to strange attractors. In order to reduce this effect we attach
to the thermostat $N_+=N_-=40$ particles from every side of the chain.

The dynamics of system (\ref{f22}) is  also completely deterministic. It should be mentioned
that it is impossible to use the initial condition (\ref{f19}) corresponding to ground state of
the system (it is stationary point of system (\ref{f22})).  We take the initial condition
         \begin{equation}
         \label{f23}
         u_n=u_0,~~~{u_n}'(0)=4(\xi_n-\frac12)\sqrt{(T_++T_-)/2},
         \end{equation}
where $\xi_n$ -- independent realizations of the random variable over the interval [0,1].

We choose $\epsilon=1$ ($g=1$), $T_+=3.05$, $T_-=2.95$, $N=80$ and integrate system (\ref{f22})
numerically with initial condition (\ref{f23}). The distribution of heat fluxes $J_n$ and local
temperatures $T_n$ is presented at Fig. \ref{fig17} (for the sake of comparison we present also
the results received bu using LT - thin lines). Within the left thermostat the heat flux grows
linearly and within the other thermostat it decreases linearly with $n$. At central part of the
chain the value of the heat flux does not depend on $n$. Linear temperature profile is formed
and the heat conductivity coefficient may be computed according to (\ref{f12}) -- $\kappa
(N)=18.4$. Use of LT gives $\kappa (N)=18.5$ (see above), i.e. the value of $\kappa$ does not
depend on the type of the thermostat.

In addition, it is possible to conclude from Fig. \ref{fig18} that the frequency distribution
of the energy of vibrations also does not depend on the type of thermostat used. It means that
for the case of the temperatures close to the value of the potential barrier the choices of NHT
or LT bring about equivalent results

The situation is strikingly different if the temperature is lower and the chain is closer to
the linear case. The Nose-Hoover thermostat is not effective in this case. In order to
illustrate this fact we use the model of harmonic chain. As it is clear from Fig. \ref{fig3}
NHT gives values of the heat flow substantially different from the correct values; at the same
times the use of LT secures much better results. That is why in the present paper we used more
complicated and consuming LT.

It should be mentioned that sometimes due to its simplicity NHT is used incorporate with LT
\cite{p23} (LT is used for the parameters of the model where NHT is not acceptable).

\end{document}